%% file: main.tex
\documentclass{style/nseJournal}
\usepackage{bm}
\usepackage{mathtools}
\usepackage{algorithm}
\usepackage{algpseudocode}
\usepackage{placeins}
\usepackage{makecell}
\usepackage{booktabs}
\usepackage{tabularx}
\usepackage{todonotes}
\usepackage[acronym,toc]{glossaries}
\loadglsentries{acronyms}

\DeclarePairedDelimiter{\inner}{\langle}{\rangle}
\begin{document}

\title{Verification and Performance Assessment of NuDEAL, a GPU-accelerated Deterministic Transport Framework on Unstructured Meshes} 

\addAuthor{Kyung Min Kim}{1,†}
\addAuthor{Jaeuk Im}{1,†}
\addAuthor{Han Gyu Lee}{2}
\addAuthor{\correspondingAuthor{Yeon Sang Jung}}{3}
\correspondingEmail{yjung@kaeri.re.kr}

\addAffiliation{1}{Seoul National University\\ 1 Gwanak-ro, Gwanak-gu, Seoul 08826, Republic of Korea}
\addAffiliation{2}{University of Michigan\\ 2355 Bonisteel Blvd, Ann Arbor, MI 48109, USA}
\addAffiliation{3}{Korea Atomic Energy Research Institute\\ 111 Daedeok-daero 989 beon-gil, Yuseong-gu, Daejeon 34057, Republic of Korea}

\titlePage

\addKeyword{Advanced reactor}
\addKeyword{Unstructured mesh}
\addKeyword{GPU acceleration}
\addKeyword{Method of Characteristics}
\addKeyword{Discrete ordinate method}

\vspace*{-0.8em}
\noindent\begingroup
\small
\textsuperscript{†}\,Current affiliation: Korea Atomic Energy Research Institute, 111 Daedeok-daero, 989 beon-gil, Yuseong-gu, Daejeon 34057, Republic of Korea
\par\endgroup

\begin{abstract}
  High-fidelity neutronic analyses of advanced reactors require deterministic transport solvers capable of handling complex unstructured geometries while maintaining computational efficiency.
  This work presents the development and verification of three GPU-accelerated deterministic solvers implemented within a unified framework, Neutronics using Deterministic Finite Element Algorithm (NuDEAL): the planar Method of Characteristics coupled with the Hybrid Finite Element Method (MOC/HFEM), the Discontinuous Galerkin Method of Characteristics (DGMOC), and the Discontinuous Finite Element discrete ordinate method (DFEM-SN).
  These solvers provide complementary capabilities for consistently solving the multigroup transport equation and can be selectively employed to balance accuracy, computational cost, and memory requirements for a given problem.
  All methods emphasize efficient GPU execution by leveraging memory alignment, compressed-flux storage, and sequential azimuthal sweeps.
  The solvers are validated on the C5G7 benchmark and applied to advanced reactor problems, including the ABTR, Empire microreactor, and MSRE.
  DFEM-SN achieved the highest accuracy, with eigenvalue errors below 50 pcm, while MOC/HFEM and DGMOC provided superior efficiency, with single-GPU runtimes comparable to those of large CPU clusters.
  The results demonstrate that deterministic GPU solvers on unstructured meshes can deliver both accuracy and scalability, enabling practical whole-core simulations for heterogeneous advanced reactors.
  The unified NuDEAL framework establishes a foundation for future extensions toward transient and multiphysics analyses on large-scale GPU architectures.
\end{abstract}

\input{sections/intro}
\input{sections/formulation}
\input{sections/implementation}
\input{sections/result}
\input{sections/conclusion}

\pagebreak
\bibliographystyle{style/ans_js}                                                                           
\bibliography{refs}

\end{document}

%% file: acronyms.tex
\newacronym{MOOSE}{MOOSE}{Multiphysics Object-Oriented Simulation Environment}
\newacronym{INL}{INL}{Idaho National Laboratory}
\newacronym{ANL}{ANL}{Argonne National Laboratory}
\newacronym{DOE-NE}{DOE-NE}{U.S. Department of Energy-Office of Nuclear Energy}
\newacronym{NEAMS}{NEAMS}{Nuclear Energy Advanced Modeling and Simulation}
\newacronym{FEM}{FEM}{Finite Element Method}
\newacronym{SN}{SN}{discrete ordinates}
\newacronym{PN}{PN}{spherical harmonics expansion}
\newacronym{SPN}{SPN}{simplified spherical harmonics expansion}
\newacronym{PBR}{PBR}{Pebble Bed Reactor}
\newacronym{MSR}{MSR}{Molten Salt Reactor}
\newacronym{FR}{FR}{Fast Reactor}
\newacronym{SFR}{SFR}{Sodium-cooled Fast Reactor}
\newacronym{HTR}{HTR}{High-Temperature Reactor}
\newacronym{LFR}{LFR}{Lead-cooled Fast Reactor}
\newacronym{FHR}{FHR}{Fluoride High-temperature Reactor}
\newacronym{LWR}{LWR}{Light Water Reactor}
\newacronym{SQA}{SQA}{Software Quality Assurance}
\newacronym{NQA-1}{NQA-1}{Nuclear Quality Assurance}
\newacronym{VTB}{VTB}{Virtual Test Bed}
\newacronym{TRISO}{TRISO}{TRi-structural ISOtropic}
\newacronym{ASME}{ASME}{American Society of Mechanical Engineers}
\newacronym{1D}{1D}{one-dimension}
\newacronym{2D}{2D}{two-dimension}
\newacronym{3D}{3D}{three-dimension}
\newacronym{PJFNK}{PJFNK}{Preconditioned Jacobian-Free Newton-Krylov}
\newacronym{PETSc}{PETSc}{Portable, Extensible Toolkit for Scientific Computation}
\newacronym{SLEPc}{SLEPc}{Scalable Library for Eigenvalue Problem Computations}
\newacronym{BlueCRAB}{BlueCRAB}{Comprehensive Reactor Analysis Bundle}
\newacronym{NCRC}{NCRC}{INL’s Nuclear Computational Resource Center}
\newacronym{CFEM}{CFEM}{Continuous Finite Element Methods}
\newacronym{DFEM}{DFEM}{Discontinuous Finite Element Methods}
\newacronym{HFEM}{HFEM}{Hybrid Finite Element Methods}
\newacronym{SSAPI}{SSAPI}{Self-Shielding Application Programming Interface}
\newacronym{CIVET}{CIVET}{Continuous Integration, Verification, Enhancement, and Testing}
\newacronym{MooseDocs}{MooseDocs}{MOOSE documentation system}
\newacronym{DG}{DG}{Discontinuous Galerkin}
\newacronym{SAAF}{SAAF}{Self-Adjoint Angular Flux formulation}
\newacronym{LS}{LS}{Least Square formulation}
\newacronym{NDA}{NDA}{Nonlinear Diffusion Acceleration}
\newacronym{LDA}{LDA}{Linear Diffusion Acceleration}
\newacronym{CMFD}{CMFD}{Coarse Mesh Finite Difference}
\newacronym{IQS}{IQS}{Improved Quasi-Static}
\newacronym{SPH}{SPH}{super-homogenization}
\newacronym{DF}{DF}{Discontinuity Factor}
\newacronym{DNP}{DNP}{Delayed Neutron Precursor}
\newacronym{PTT}{PTT}{Pebble Tracking Transport}
\newacronym{AMG}{AMG}{Algebraic Multi-Grid}
\newacronym{KBA}{KBA}{Koch-Baker-Alcouffe}
\newacronym{FIFO}{FIFO}{First-In-First-Out}
\newacronym{DoFs}{DoFs}{Degrees of Freedom}
\newacronym{GMRes}{GMRes}{Generalized Minimal Residual method}
\newacronym{VNM}{VNM}{Variational Nodal Method}
\newacronym{XML}{XML}{Extensible Markup Language}
\newacronym{ADF}{ADF}{Assembly Discontinuity Factors}
\newacronym{EDC}{EDC}{Equivalence Dancoff-factor Cell}
\newacronym{ILSS}{ILSS}{Iterative Local Spatial Self-shielding}
\newacronym{DTL}{DTL}{Decay Transmutation Library}
\newacronym{ENDF}{ENDF}{Evaluated Nuclear Data File}
\newacronym{ADM}{ADM}{Adding and Doubling Method}
\newacronym{CRAM}{CRAM}{Chebyshev Rational Approximation Method}
\newacronym{GS}{GS}{Gauss-Seidel}
\newacronym{SGE}{SGE}{sparse Gaussian elimination}
\newacronym{RGMB}{RGMB}{Reactor Geometry Mesh Builder}
\newacronym{ABTR}{ABTR}{Advanced Burner Test Reactor}
\newacronym{RMS}{RMS}{Root Mean Square}
\newacronym{SZ}{SZ}{Spectrum Zone}
\newacronym{PDT}{PDT}{Parallel Deterministic Transport}
\newacronym{MSRE}{MSRE}{Molten-Salt Reactor Experiment}
\newacronym{MSFR}{MSFR}{Molten Salt Fast Reactor}
\newacronym{TREAT}{TREAT}{Transient REActor Test}
\newacronym{HTTR}{HTTR}{High-Temperature engineering Test Reactor}
\newacronym{GC}{GC}{Gas-Cooled}
\newacronym{gFHR}{gFHR}{generic Fluoride salt-cooled High-temperature Reactor}
\newacronym{NTP}{NTP}{Nuclear Thermal Propulsion}
\newacronym{gHPMR}{gHPMR}{generic High-Pressure Microplate Reactor}
\newacronym{gPBR}{gPBR}{generic Pebble Bed Reactor}
\newacronym{NRC}{NRC}{Nuclear Regulatory Commission}
\newacronym{ART}{ART}{Advanced Reactor Technology}
\newacronym{NRIC}{NRIC}{National Reactor Innovation Center}
\newacronym{ATR}{ATR}{Advanced Test Reactor}
\newacronym{DOF}{DoF}{Degree of Freedom}

\newacronym{DGMOC}{DGMOC}{Discontinuous Galerkin Method of Characteristics}
\newacronym{MOC}{MOC}{Method of Characteristics}
\newacronym{MC}{MC}{Monte Carlo}
\newacronym{DFEMSN}{DFEM-SN}{Discontinuous Finite Element Method with discrete ordinate}
\newacronym{NuDEAL}{NuDEAL}{Neutronics using Deterministic Finite
Element Algorithms}
\newacronym{XS}{XS}{cross section}
\newacronym{FSR}{FSR}{flat source region}
\newacronym{LLNL}{LLNL}{Lawrence Livermore National Laboratory}
\newacronym{STRAUM}{STRAUM}{SN Transport for Radiation Analysis with Unstructured Meshes}
\newacronym{GCMFD}{GCMFD}{Generalized Coarse Mesh Finite Difference}
\newacronym{SIMD}{SIMD}{Single Instruction Multiple Data}
\newacronym{GPU}{GPU}{Graphics Processing Unit}
\newacronym{CUDA}{CUDA}{Compute Unified Device Architecture}
\newacronym{VRAM}{VRAM}{Video Random Access Memory}
\newacronym{DoF}{DoF}{Degree of Freedom}
\newacronym{HEDS}{HEDS}{half-edge data structure}
\newacronym{PWR}{PWR}{pressurized water reactor}
\newacronym{PRAGMA}{PRAGMA}{Power Reactor Analysis using GPU-based Monte Carlo Algorithm}
\newacronym{LANL}{LANL}{Los Alamos National Laboratory}
\newacronym{ORNL}{ORNL}{Oak Ridge National Laboratory}
\newacronym{RB}{RB}{red-black}

%% file: sections/intro.tex
\section{Introduction}
\label{sec:intro}

Various reactor concepts, including molten salt reactors, liquid-metal-cooled fast reactors, and heat-pipe microreactors, are being investigated for advanced reactor design studies. Conventional analysis methods tailored to specific reactor types have become less effective due to strong heterogeneity effects and non-conventional core configurations in these reactor designs. There is growing interest in alternative reactor analysis methods capable of high-fidelity core analysis for diverse advanced reactor designs. These methods should feature two essential capabilities: neutron transport theory and flexible representation of core geometry. The continuous-energy \gls{MC} neutron transport method is an ideal candidate.
It is widely used for advanced reactor analyses, but it is unlikely to become mainstream in the future due to its substantial computational requirements, particularly for transient and depletion analyses.

Over the past decades, several CPU-based deterministic transport solvers have advanced in their ability to handle unstructured geometries. PROTEUS-MOC, developed at the \gls{ANL}, adopted \gls{DGMOC} that extends the \gls{MOC} over extruded 3D domains through a discontinuous Galerkin formulation in the axial direction, demonstrating accurate 3D capabilities~\cite{marin-laflechePROTEUSMOC3DDETERMINISTIC2013}.Within the \gls{MOOSE} framework~\cite{permannMOOSEEnablingMassively2020}, Rattlesnake~\cite{wangRattlesnakeMOOSEBasedMultiphysics2021} and its successor Griffin~\cite{wangGriffinMOOSEbasedReactor2025} implemented finite-element \gls{SN} solvers, including \gls{DFEMSN}, capable of operating on arbitrary polygonal or polyhedral meshes with tight coupling to multiphysics modules.
These efforts to utilize unstructured meshes established the numerical and parallel foundations for deterministic transport on general meshes.
Yet, they rely on large-scale CPU clusters and distributed-memory parallelism, making them prohibitive for ordinary researchers and institutions that want to employ unstructured meshes for detailed design.

Recent studies have investigated \gls{GPU} acceleration for deterministic transport.
One of the earliest GPU-based implementations of \gls{MOC} ported a 2D solver to \gls{CUDA}, achieving over a hundredfold speedup on a single GPU~\cite{zhangAcceleratingThreedimensionalMOC2013}.
This work demonstrated that \gls{MOC} can be efficiently mapped onto massively parallel architectures, albeit with limited support for structured geometries.
OpenMOC, an open-source \gls{MOC} code, subsequently incorporated GPU acceleration features, focusing on vectorized Jacobi sweeps and optimized memory layouts for flat-source \gls{MOC} calculations~\cite{boydOpenMOCMethodCharacteristics2014}.
While OpenMOC demonstrated substantial speedups and provided an accessible platform for GPU algorithm development, its GPU implementation primarily targeted moderately sized pin-cell or lattice problems rather than large whole-core configurations.
More recently, the planar \gls{MOC} based transport code, nTRACER~\cite{jungPracticalNumericalReactor2013}, integrated GPU kernels into its production-grade 2D/1D direct whole-core calculation framework.
Within this framework, the \gls{MOC} data layout was restructured, Jacobi energy sweeping was adopted to enhance \gls{SIMD} vectorization, mixed-precision arithmetic was introduced, and CPU–GPU concurrency was exploited to overlap source updates with ray tracing~\cite{choiPracticalAccelerationDirect2021}.
The results indicate that a single consumer-grade RTX-class GPU can deliver ray-tracing performance equivalent to hundreds of CPU cores, enabling full 3D whole-core transport calculations within minutes on affordable clusters.
However, the GPU implementation relies on structured radial discretizations.
Together, these efforts demonstrate the viability of GPU acceleration for deterministic \gls{MOC}, but they generally focus on structured geometries and/or 2D/1D synthetic approaches.

Several studies have exploited GPUs to accelerate \gls{SN} methods.
One of the earliest demonstrations of GPU-based \gls{SN} sweeps on unstructured meshes employed a predetermined sweep order, enabling parallel execution of all energy groups on a single GPU~\cite{gongGPUAcceleratedSimulations2011}.
Although their method achieved more than an order-of-magnitude speedup for 2D problems, it remained limited to relatively small meshes and did not address full 3D sweep dependencies.
Subsequent work at \gls{LLNL} ported the 3D sweep algorithm of the structured-mesh \gls{SN} code Ardra~\cite{hanebutteARDRAScalableParallel1999} to \gls{CUDA}, adopting a block-synchronous strategy that delegated sweep-ordering to thread blocks~\cite{kunenPorting3DDiscrete2019}.
This study demonstrated the feasibility of GPU offloading for production-scale \gls{SN} kernels.
Still, the port targeted only the sweeping operator and did not extend to full multigroup solvers or unstructured grids.
More recently, \gls{STRAUM}, an unstructured-mesh \gls{SN} code, incorporated a fully GPU-accelerated sweep kernel together with a GPU-resident multigroup Krylov solver~\cite{zhangGPUbasedMultigroupDiscrete2025}.
The algorithm exploits parallelism across energy groups, angles, and cells while strictly enforcing sweep dependencies through intra-block synchronization.
\gls{STRAUM} demonstrated speedups of 70–200\(\times\) on modern GPUs and successfully scaled to multi-GPU configurations with more than 40\% memory reduction per device.
Nevertheless, \gls{STRAUM} only focuses on the parallelization and does not consider the prohibitive amount of angular flux storage that occupies GPU memory during a sweep.
These efforts in GPU-based \gls{SN} solvers collectively establish important foundations for GPU-accelerated deterministic transport algorithms.
However, in common, they focused on \gls{SN} sweeps, either on structured or unstructured meshes, rather than on efficient memory management of a large number of angular flux variables within the limited GPU \gls{VRAM}.

To address the gap described above, this work presents efficient GPU parallelization methods for three complementary neutron transport solvers implemented in a unified, GPU-accelerated code system, \gls{NuDEAL}, designed to leverage unstructured meshes for advanced reactor analysis.
The first is the planar \gls{MOC} combined with the 3D \gls{HFEM}.
This method resembles conventional direct whole-core calculation schemes, which combine the planar \gls{MOC} and the axial nodal method within the framework of \gls{CMFD}.
\gls{HFEM} solves the 3D coarse domain and serves as an axial leakage provider to the planar \gls{MOC}.
This scheme eliminates potential inaccuracies and instabilities arising from the transverse-leakage approximation used in the axial nodal solver.
Another is the \gls{DGMOC} solver, which efficiently solves the transport equation directly in 3D.
These two solver options are limited to axially extruded core models.
The last, \gls{DFEMSN}, is free from such limitations in core geometry representations at the expense of extreme memory requirements.
Across the three solvers, several GPU-oriented strategies were adopted to ensure high throughput on unstructured meshes.
This enables the solution of the neutron transport equation in the presence of structural deformation.
Each solver incorporates method-specific strategies suited to its numerical structure.
The planar \gls{MOC}/\gls{HFEM} solver employs a flattened, ray-centric parallelism with precomputed segment data, enabling highly efficient, independent characteristic sweeps.
The \gls{DGMOC} solver adopts a sequential azimuthal sweep to limit memory pressure while maintaining high occupancy, computing axial \gls{DG} moments plane-by-plane.
For \gls{DFEMSN}, memory compression of angular-flux buffers and \gls{DoF}-level parallelization maximize concurrency during transport sweeps.
Together, these optimizations enable \gls{NuDEAL} to maintain numerical consistency while fully exploiting the fine-grained parallelism of modern GPUs.

The remainder of the paper is organized as follows.
Section~\ref{sec:formulation} presents the methodologies of \gls{NuDEAL}, starting with multigroup discretization, and finite element weak forms of each method.
Section~\ref{sec:implementation} summarizes numerical implementation and GPU-oriented optimizations in \gls{NuDEAL}.
The numerical results are demonstrated in Section~\ref{sec:result}, which reports verification and performance analyses and discusses their implications.
Section~\ref{sec:conclusion} concludes the paper with a summary and outlook for future work.

%% file: sections/formulation.tex
\section{Deterministic Methods Employed in \gls{NuDEAL}}
\label{sec:formulation}

The \gls{NuDEAL} code solves the steady-state multigroup neutron transport equation in differential form, given by
\begin{align}
  \mathbf{\Omega} \cdot \nabla \varphi_g(\mathbf{r}, \mathbf{\Omega}) + \Sigma_{t g}(\mathbf{r}) \varphi_g(\mathbf{r}, \mathbf{\Omega}) & = \sum_{g'=1}^{G} \int_{4\pi} \Sigma_{s g' \to g}(\mathbf{r}, \mathbf{\Omega}' \to \mathbf{\Omega}) \varphi_{g'}(\mathbf{r}, \mathbf{\Omega}') d \mathbf{\Omega}' \notag \\ & + \frac{\chi_g(\mathbf{r})}{k_{eff}} \sum_{g'=1}^{G} \nu \Sigma_{f g'}(\mathbf{r}) \phi_{g'}(\mathbf{r}),
  \label{eqn:MGNTE}
\end{align}
where the terms in Eq.~(\ref{eqn:MGNTE}) follows standard notation.

To obtain a tractable system, the scattering source is commonly expanded in Legendre polynomials as
\begin{align}
  \Sigma_{s g' \to g}(\mathbf{r}, \mathbf{\Omega}' \to \mathbf{\Omega}) & =
  \sum_{l=0}^{L} \frac{2l + 1}{4\pi} P_l(\mathbf{\Omega} \cdot \mathbf{\Omega}') \Sigma_{sl g' \to g}(\mathbf{r}),
\end{align}
where $P_{l}$ and $ \Sigma_{sl g' \to g}(\mathbf{r})$ are the l-th order Legendre polynomial and the corresponding higher order scattering \gls{XS}. 
This allows isotropic or anisotropic scattering approximations depending on the truncation order \(L\).
The spatial domain \(\mathcal{D}\) is discretized into non-overlapping polyhedral elements \(\mathcal{D}_e\) such that \(\mathcal{D} = \bigcup_{e} \mathcal{D}_e\).
Eq.~(\ref{eqn:MGNTE}) is redefined over each element \(\mathcal{D}_e\):
\begin{align}
  \mathbf{\Omega} \cdot \nabla \varphi_{ge}(\mathbf{r}, \mathbf{\Omega}) +
  \Sigma_{tge}(\mathbf{r}) \varphi_{ge}(\mathbf{r}, \mathbf{\Omega}) & =
  q_{sge}(\mathbf{r}, \mathbf{\Omega}) +
  q_{fge}(\mathbf{r}),
  \label{eqn:MGNTE_element}
\end{align}
where $q_{sge}$ is the scattering source term and $q_{fge}$ is the fission source term for group \(g\) in element \(e\).
Eq.~(\ref{eqn:MGNTE_element}) provides a common finite-element foundation for all subsequent formulations.

\input{sections/formulation/MOC.tex}
\input{sections/formulation/DFEM-SN.tex}

%% file: sections/formulation/MOC.tex
\subsection{Planar \gls{MOC} Based Methods}
\label{sec:MOC_DGMOC}

\gls{MOC} is a widely used deterministic approach for solving the steady-state neutron transport equation in heterogeneous domains by converting the integro-differential form into a set of one-dimensional differential equations along discrete characteristic lines that span the entire spatial domain and angular space.
In this approach, the spatial domain is discretized into regions with uniform material properties and source terms.
Each such region is commonly referred to as a \gls{FSR} and generally corresponds to a single finite element in the unstructured-mesh sense.
The neutron angular flux is then integrated along discrete straight-line characteristic paths determined by the region boundaries.
For a given direction \(\mathbf{\Omega}_m\) and spatial point \(\mathbf{r}\), the steady-state multigroup transport equation can be expressed as
\begin{equation}
  \mathbf{\Omega}_m \cdot \nabla \varphi_m(\mathbf{r}) + \Sigma_t(\mathbf{r}) \varphi_m(\mathbf{r})
  = q_m(\mathbf{r}),
  \label{eqn:MOC_eq}
\end{equation}
where \(q_m(\mathbf{r})\) represents the total source, including scattering and fission contributions, and the group index is omitted for brevity.
Integrating Eq.~\eqref{eqn:MOC_eq} along a characteristic segment of length \(s\) within a \gls{FSR} yields
\begin{equation}
  \varphi_m(s) = \varphi_m(0) e^{-\Sigma_{t} s} + \frac{q_m}{\Sigma_t} \left( 1 - e^{-\Sigma_{t} s} \right),
  \label{eqn:MOC_sol}
\end{equation}
which provides a direct relation between the incoming and outgoing angular fluxes along the track. 
The scalar flux $\phi(\mathbf{r})$ is then obtained by angular integration over all discretized angular directions.

Two approaches are investigated to retain the planar \gls{MOC}'s 3D solution capability: axial \gls{DG} discretization and coarse-mesh 3D \gls{HFEM}.
The \gls{DG} discretization is directly applied to the planar \gls{MOC} equation, Eq.~\eqref{eqn:MOC_eq}, resulting in the \gls{DGMOC} method.
In this method, all angular flux assigned to the characteristic lines varies axially within an axial slab.
Therefore, \gls{DGMOC} can produce accurate 3D solutions but requires substantially more computational time than the conventional planar \gls{MOC}.
On the other hand, the coarse mesh 3D \gls{HFEM} approach retains the planar \gls{MOC} formulation in the radial plane. 
Although this approach has the drawback that the separate solver evaluates axial behavior, it significantly reduces the computational cost compared to \gls{DGMOC} while still providing an effective representation of axial leakage and global 3D effects.
It is noted that both \gls{MOC}-based methods are applicable only to axially extruded geometries.

\subsubsection{\gls{DGMOC} Formulation}

\gls{DGMOC} extends the planar \gls{MOC} by applying a \gls{DG} discretization along the axial direction.
The angular flux $\varphi_m(\mathbf{r})$ and the source $q_m(\mathbf{r})$ are expanded using $n$ basis functions $u_p^i(z)$ within each axial element $i$ as
\begin{equation}
  \varphi_{m}^i(x,y,z) = \sum_{p=1}^{n} \varphi_{mp}^i(x,y) u_p^i(z), \quad
  q_{m}^i(x,y,z) = \sum_{p=1}^{n} q_{mp}^i(x,y) u_p^i(z).
  \label{eqn:DGMOC_expansion}
\end{equation}
Multiplying Eq.~\eqref{eqn:MOC_eq} by a test function $u_q^i(z)$ and integrating over the axial element gives the weak form,
\begin{equation}
  \int_{z_{i-1}}^{z_i} u_q^i(z) \mathbf{\Omega}_m \cdot \nabla \varphi_m^i \, dz
  + \int_{z_{i-1}}^{z_i} u_q^i(z) \Sigma_t \varphi_m^i \, dz
  = \int_{z_{i-1}}^{z_i} u_q^i(z) q_m^i \, dz.
  \label{eqn:DGMOC_weak}
\end{equation}
For the directions $\mathbf{\Omega}_{m,z} > 0$, Eq. ~\eqref{eqn:DGMOC_weak} can be simplified as:
\begin{equation}
\begin{gathered}
  (\Omega_{m,x}\frac{\partial}{\partial x}+\Omega_{m,y}\frac{\partial}{\partial x})\int_{z_{i-1}}^{z_i} u_q^i(z)  \varphi_m^i(x, y,z_{i}) \, dz \\
  + \Omega_{m,z} u_q^i(z_{i+1})\varphi_m^i(x, y,z_{i})\ 
  + \int_{z_{i-1}}^{z_i} u_q^i(z) \Sigma_t \varphi_m^i \, dz \\
  = \int_{z_{i-1}}^{z_i} u_q^i(z) q_m^i \, dz.
   + \Omega_{m,z} u_q^i(z_{i})\varphi_m^i(x, y,z_{i-1})\
  \label{eqn:DGMOC_weak_up}
\end{gathered}
\end{equation}
By employing the upwinding scheme, the axial coupling appearing in the RHS can be resolved. 
Then, the angular flux along each characteristic segment is computed using an equation similar to Eq.~\eqref{eqn:MOC_sol}.
The resulting system of equations over every ray segment is a matrix exponential equation, having a size of $n \times n$, where $n$ is the number of axial basis functions used in Eq.~\eqref{eqn:DGMOC_expansion}.
In this study, the expansion order of the basis function is limited to first order to account for the complexity of matrix exponential evaluations. 
Readers are recommended to refer to~\cite{marin-laflechePROTEUSMOC3DDETERMINISTIC2013} for a comprehensive derivation of the method.
It is noted that this scheme is guaranteed to converge to the true solution of the transport equation as the spatial and angular discretizations are refined, whereas conventional planar \gls{MOC}-based schemes do not.

\subsubsection{Planar \gls{MOC}/\gls{HFEM} Formulation}

\gls{HFEM}~\cite{raviartPrimalHybridFinite1977} is also known as the \gls{VNM}~\cite{vladimirovMathematicalProblemsOneVelocity1961}\cite{dilberVariationalNodalMethods1985}, a generalized nodal method working on unstructured meshes.
In contrast to conventional nodal methods, \gls{HFEM} does not employ the transverse-integration technique. 
Instead, the intranodal solution is expanded directly over the entire three-dimensional node.
Consequently, \gls{HFEM} avoids the limitation inherent in the transverse-leakage approximation.
Moreover, unlike conventional nodal methods that rely on a fixed intranodal expansion order, the solution accuracy in \gls{HFEM} can be systematically improved by increasing the order of the intranodal expansion, which is referred to as \textit{p-refinement}.

In this study, \gls{HFEM} is applied to both the diffusion and $SP_3$ equations, serving as the coarse-mesh three-dimensional solver within the planar \gls{MOC}/\gls{HFEM} method.
Planar \gls{MOC} provides high-fidelity radial transport on a fine unstructured mesh, and \gls{HFEM} supplies the axial coupling on a much coarser mesh.
This division of roles allows the hybrid method to retain the accuracy of \gls{MOC} in the radial direction while obtaining efficient 3D solutions through \gls{HFEM}.
\gls{HFEM} also provides the surface currents required for \gls{CMFD} acceleration, and these currents are combined with the radial \gls{MOC} currents to enforce consistency over the whole core.
Through this workflow, \gls{HFEM} plays a central role in extending planar \gls{MOC} to full 3D whole-core calculations without incurring the prohibitive computational cost.

Derivation of \gls{HFEM} begins with the second-order neutron transport equation and applies a hybrid finite-element approach to discretize the spatial domain.
Although any angular approximations can be applied, such as the $S_N$~\cite{sunDiscreteordinatesVariationalNodal2024} or $P_N$~\cite{dilberVariationalNodalMethods1985} method, the diffusion form is presented here for brevity.
Assuming the void boundary condition is not imposed, the functional for the diffusion \gls{HFEM} is given as
\begin{equation}
  \mathcal{F}(\phi, J) = \sum_{e} \left[ \int_{\mathcal{D}_{e}} D (\nabla \phi)^2 dV + \int_{\mathcal{D}_{e}} \Sigma_a \phi^2 dV - 2 \int_{\mathcal{D}_{e}} Q \phi dV \right] + 2 \sum_{f} \int_{\partial \mathcal{D}_{e}} J \phi dS,
  \label{eqn:HFEM_functional}
\end{equation}
where \(\phi\) is the scalar flux, \(J\) is the net current across element faces, \(D\) is the diffusion coefficient, \(\Sigma_a\) is the absorption \gls{XS}, and \(Q\) is the source term. Eq.~\eqref{eqn:HFEM_functional} is minimized with respect to both \(\phi\) and \(J\) to derive the element-wise weak form:
\begin{equation}
  \left( \nabla \phi^{*}, D \nabla \phi \right)_{\mathcal{D}_{e}} +
  \left( \phi^{*}, \Sigma_a \phi \right)_{\mathcal{D}_{e}} +
  \inner{\phi^{*}, J}_{\partial \mathcal{D}_{e}} +
  \inner{J^{*}, \phi}_{\partial \mathcal{D}_{e}} =
  \left( \phi^{*}, Q \right)_{\mathcal{D}_{e}},
  \label{eqn:HFEM_weak_form}
\end{equation}
where \((\cdot, \cdot)_{\mathcal{D}_{e}}\) denotes the volume inner product over element \(\mathcal{D}_{e}\), and \(\inner{\cdot, \cdot}_{\partial \mathcal{D}_{e}}\) denotes the surface inner product over the element boundary \(\partial \mathcal{D}_{e}\).
For the spatial discretization, the scalar flux \(\phi\) is approximated by an orthogonal polynomial expansion within each element, while the net current \(J\) is represented using basis orthogonal functions defined on the element faces.
The scalar flux and the neutron current are expressed as
\begin{equation}
  \phi_{e}(\mathbf{r}) = \sum_{i=1}^{N_p} \phi_{ei} u_i(\mathbf{r}),
  \label{eqn:HFEM_flux_expansion}
\end{equation}
\begin{equation}
  J_{f}(\mathbf{r}) = \sum_{j=1}^{N_f} J_{efj} v_j(\mathbf{r}),
  \label{eqn:HFEM_current_expansion}
\end{equation}
where \(u_i(\mathbf{r})\) and \(v_j(\mathbf{r})\) are the orthogonal polynomial basis functions on the volume $e$ and the face $f$ respectivley, and \(N_p\) and \(N_f\) is the number of used basis functions.
Substituting Eqs.~\eqref{eqn:HFEM_flux_expansion} and~\eqref{eqn:HFEM_current_expansion} into Eq.~\eqref{eqn:HFEM_weak_form} leads to a system of algebraic equations for the coefficients \(\phi_{e}\) and \(J_{f}\).
The resulting algebraic equations are solved by recasting them into a response matrix form expressed in terms of partial currents:

\begin{equation}
  \mathbf{J^{+}_e}=\mathbf{R}_e \mathbf{J^{-}_e} + \mathbf{B}_e \mathbf{Q_e} ,
  \label{eqn:HFEM_flux_response}
\end{equation}
where $\mathbf{J}^{\pm}_e$ is the outgoing and incoming partial currents including entire faces in the element $\mathcal{D}_{e}$.
This response matrix form can be efficiently solved using a \gls{RB} iterative solution algorithm, in which the elements are partitioned into two disjoint sets such that no element in one set directly couples with another element in the same set.
As a result, all elements belonging to the same color can be solved independently and simultaneously, making the RB scheme particularly well-suited for parallel execution.
This property enables element-wise local solves without global synchronization within each color, thereby improving parallel efficiency.
It is recommended to refer to~\cite{zhangVariationalNodalMethods2022} and~\cite{smithDIF3DVARIANT110Decade2014} for detailed derivations.
The corresponding derivation for the $SP_3$ formulation can be straightforwardly obtained by extending the diffusion formulation.

\gls{MOC} and \gls{HFEM} methods are coupled to leverage the strengths of both approaches.
\gls{MOC} is employed to solve the transport equation on a 2D fine mesh, while \gls{HFEM} is performed to solve diffusion or low-order angular equations on a 3D coarse mesh.
Unfortunately, full consistency between them is not guaranteed since the two methods operate on different spatial discretizations~\cite{kim2D3DGPUaccelerated2024}.
Therefore, \gls{CMFD} acceleration is mediated to the \gls{MOC}/\gls{HFEM} method to enforce global neutron balance consistency, which is summarized in Algorithm~\ref{alg:MOCHFEM_CMFD}.
\begin{algorithm}[htbp]
  \caption{MOC/HFEM hybrid method with CMFD acceleration}
  \label{alg:MOCHFEM_CMFD}
  \begin{algorithmic}[1]
    \Repeat
    \State Perform MOC on the fine mesh
    \State Homogenize fine-mesh MOC solution onto coarse mesh for HFEM
    \State Solve HFEM on the coarse mesh
    \State Solve CMFD on the coarse mesh.
    \State \quad Use MOC and HFEM currents for radial and axial surface, respectively
    \State Update fine-mesh MOC solution with accelerated results
    \Until{convergence}
  \end{algorithmic}
\end{algorithm}
\FloatBarrier

%% file: sections/formulation/DFEM-SN.tex
\subsection{\gls{DFEMSN}}
\label{sec:DFEM_SN}

\gls{DFEMSN} solves the multigroup neutron transport equation by spatially discretizing the domain with discontinuous finite elements and by discretizing the angular variable using the discrete ordinates (\(S_N\)) method. 
In \acrshort{DFEM}, the computational domain is divided into finite elements $\mathcal{D}_e$, within which the angular flux is approximated using a set of local basis functions $u_i(\mathbf{r})$:
\begin{equation}
  \varphi_{em}(\mathbf{r}) = \sum_{i=1}^{N_p} \varphi_{emi} u_i(\mathbf{r}), \quad
  q_{em}(\mathbf{r}) = \sum_{i=1}^{N_p} q_{emi} u_i(\mathbf{r}),
  \label{eqn:DFEM_basis}
\end{equation}
where $N_p$ denotes the number of basis functions per element.
Multiplying Eq.~\eqref{eqn:MOC_eq} by a test function $u_j(\mathbf{r})$ and integrating over the element yields the weak form
\begin{equation}
  \int_{\mathcal{D}_e} u_j \, \mathbf{\Omega}_m \cdot \nabla \varphi_{em} \, dV
  + \int_{\mathcal{D}_e} u_j \Sigma_t \varphi_{em} \, dV
  = \int_{\mathcal{D}_e} u_j q_{em} \, dV.
  \label{eqn:DFEM_weak}
\end{equation}
Applying integration by parts to the streaming term gives
\begin{align}
  -\int_{\mathcal{D}_e} (\nabla u_j) \cdot \mathbf{\Omega}_m \varphi_{em} \, dV
  + \int_{\partial \mathcal{D}_e} u_j \varphi_{em} \, \mathbf{\Omega}_m \cdot \mathbf{n} \, dS
  + \int_{\mathcal{D}_e} u_j \Sigma_t \varphi_{em} \, dV
  = \int_{\mathcal{D}_e} u_j q_{em} \, dV,
  \label{eqn:DFEM_parts}
\end{align}
where $\mathbf{n}$ is the unit outward normal on the element surface $\partial \mathcal{D}_e$.
An upwind numerical flux is applied to define the interface angular flux as
\begin{equation}
  \varphi_m^{*} =
  \begin{cases}
    \varphi_m^{\text{up}},   & \text{if } \mathbf{\Omega}_m \cdot \mathbf{n} > 0, \\
    \varphi_m^{\text{down}}, & \text{otherwise},
  \end{cases}
  \label{eqn:DFEM_upwind}
\end{equation}
which ensures numerical stability and physical consistency in the transport of particles across element boundaries.
Detailed derivations of the algebraic matrix form and element-wise basis functions for the weak form Eq.~\eqref{eqn:DFEM_weak} can be found in~\cite{wangRattlesnakeTheoryManual2018}.

%% file: sections/implementation.tex
\section{Implementation and GPU Parallelization Strategies}
\label{sec:implementation}

The three complementary methods implemented within the \gls{NuDEAL} framework leverage GPU architectures to achieve high computational performance for large-scale neutron transport problems on unstructured meshes.
This section outlines the key aspects of the implementation, including data structures, parallelization strategies, and optimization techniques that exploit GPU parallelism to maximize utilization across the methodologies.

\input{sections/implementation/MOCHFEM}
\input{sections/implementation/DGMOC}
\input{sections/implementation/DFEM-SN}

%% file: sections/implementation/MOCHFEM.tex
\subsection{Fine-mesh Planar \gls{MOC} with Coarse Mesh \gls{HFEM}}
\label{sec:MOCHFEM}

Unlike in traditional structured regular grids, unstructured meshes are not assumed to exhibit any pattern or regularity in their element arrangement.
Therefore, the \gls{MOC} implementation must account for the connectivity of each element to construct ray segment data before performing ray tracing.
For efficient representation of 2D meshes, \gls{NuDEAL} uses a \gls{HEDS}~\cite{botschOpenMeshaGenericEfficient1999}, which enables efficient traversal and navigation of mesh elements in 2D geometry.
Ray generation is also distinct from structured-mesh \gls{MOC}, which traditionally employs a modular ray approach for Cartesian and hexagonal grids.
\gls{NuDEAL} supports two types of ray generation methods:
\begin{enumerate}
  \item Modular ray approach~\cite{kochunasHybridParallelAlgorithm2013}, not for the pin-wise or assembly-wise repeated structures, but for the bounding box of the whole domain; and
  \item Back projection method~\cite{palmiottiUNICUltimateNeutronic2007}, which generates rays from the boundary into the domain based on specified angular quadrature sets.
\end{enumerate}

The ray-mesh intersection is then computed and established once the rays have been cast.
This procedure computes the segment length of each ray that passes through the polygonal elements and identifies the corresponding finite element as the ray reaches the mesh boundary.
The algorithm for ray-mesh intersection has been developed using \gls{HEDS} traversal to efficiently navigate the mesh~\cite{kimGPUacceleratedMethodCharacteristics2023}.

\begin{figure}[htbp]
  \centering
  \includegraphics[width=0.8\textwidth]{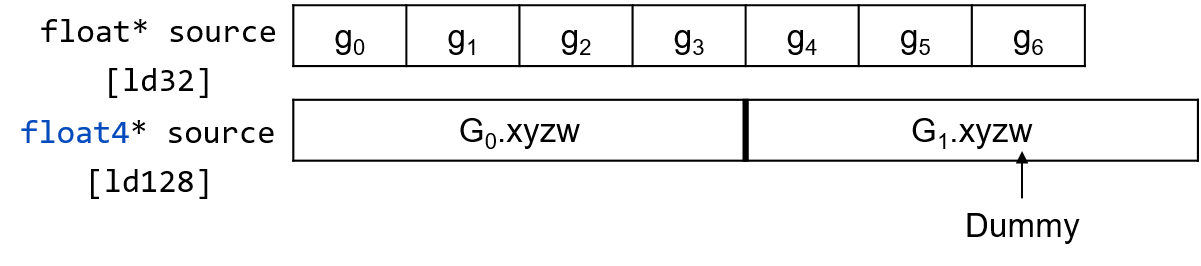}
  \caption{32-bit (top) and 128-bit aligned (bottom) memory layout.}
  \label{fig:MOC_memalign}
\end{figure}

GPU acceleration of ray tracing on unstructured grids follows similar strategies as those used in structured \gls{MOC} implementations, carefully described in~\cite{choiPracticalAccelerationDirect2021}.
On unstructured meshes, irregular data access patterns are unavoidable due to the lack of regular connectivity, thereby degrading GPU performance.
To mitigate this effect, memory alignment is applied to the cross-section (\gls{XS}), flux, and source data, as illustrated in Fig.~\ref{fig:MOC_memalign}.
Efficient memory access is critical for achieving high performance on GPUs, and 128-bit aligned memory access patterns are particularly effective in this regard.
Accordingly, the \gls{XS}, flux, and source data for each element are stored in contiguous memory locations and packed in groups of four energy groups to match the 128-bit alignment.
When the number of energy groups is not a multiple of four, additional alignment entries are introduced to maintain the 128-bit alignment, as illustrated in Fig.~\ref{fig:MOC_memalign}.
This layout enables the GPU to load data with a single 128-bit memory transaction rather than multiple smaller transactions, thereby improving memory access efficiency and overall performance.
\FloatBarrier

\begin{figure}[htbp]
  \centering
  \includegraphics[width=0.9\textwidth]{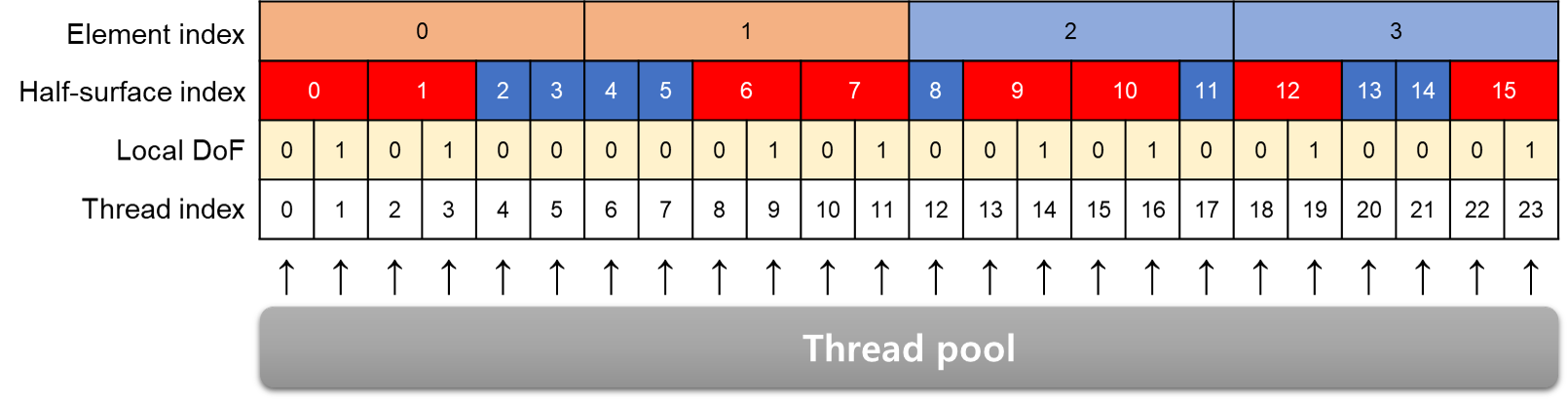}
  \caption{Data serialization example for RB sweep in HFEM solution procedure.}
  \label{fig:HFEM_dataserial}
\end{figure}

\begin{figure}[htbp]
  \centering
  \includegraphics[width=0.7\textwidth]{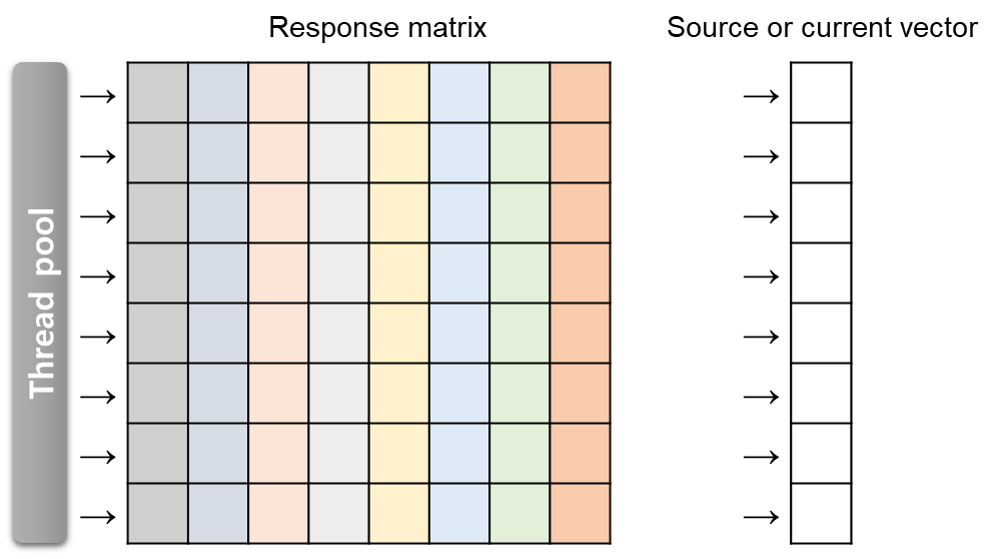}
  \caption{GPU thread access pattern to the response matrix and operand vectors during RB sweep.}
  \label{fig:HFEM_accesspattern}
\end{figure}

The \gls{HFEM} calculation procedure comprises two parts: the preassembling phase and the solution phase. 
The preassembling includes two matrix constructions and corresponding matrix inversions for the reaction and response matrix, respectively. 
Two matrix inversions, for the reaction matrix inversion and the subsequent response matrix, are performed using the cuBLAS library~\cite{nvidiaNVBLASLibrary2020} provided by the CUDA toolkit. 
Specifically, \texttt{cublasDgetrfBatched} is employed to factorize a batch of dense matrices into lower and upper triangular matrices, and then \texttt{cublasDgetriBatched} is called to do matrix inversions with those triangular matrices. 
The solution phase consists of three procedures: source update, \gls{RB} sweep, and flux reconstruction. 
The most time-consuming part of the computation is the \gls{RB} sweep, which is invoked repeatedly and involves a large number of small matrix--vector multiplications per call, necessitating careful optimization.
To this end, each GPU thread is assigned to process a single \gls{DoF}, corresponding to an undetermined coefficient of a basis function, within a local system for each interface.
Each thread has unique data corresponding to the element index, the half-surface index, and the local \gls{DoF} order, all of which are serialized in advance, as illustrated in the example in Fig.~\ref{fig:HFEM_dataserial}. 
Note that the interface polynomial expansion order may differ across the entire domain.
In the actual implementation, two matrix--vector multiplications are carried out using the precomputed response matrices: one for the surface source term and the other for the incoming partial current, as given in Eq.~\eqref{eqn:HFEM_flux_response}.
Each matrix--vector operation is performed column by column, with the resulting vector updated concurrently.
Because each column of the response matrix is stored contiguously in memory, multiple threads can efficiently access the data during the matrix--vector operation, as illustrated in Fig.~\ref{fig:HFEM_accesspattern}.

\FloatBarrier

%% file: sections/implementation/DGMOC.tex
\subsection{Sequential Azimuthal Angle Sweep in \gls{DGMOC}}
\label{sec:DGMOC_implementation}

The \gls{DGMOC} module shares the ray tracing and segment data structures described in the \gls{MOC} implementation.
The planar \gls{MOC} tracks all rays cast over the entire set of discrete angles in case of the isotropic scattering source to maximize GPU throughput.
However, this approach cannot be directly applied to \gls{DGMOC}, since the axial upwinding scheme requires the angular flux to be stored only on the upwind interface plane, which introduces an explicit dependence of the source term on the discrete ordinates.
Therefore, several algorithmic and data-structure adaptations are introduced to prevent excessive memory usage for storing the angular source at the interface between neighboring axial planes.
To accommodate this, \gls{NuDEAL} employs a sequential azimuthal angle sweep strategy, as outlined in Algorithm~\ref{alg:DGMOC_sweep}.
The entire azimuthal angle set is partitioned into smaller-angle blocks, or into a single azimuthal angle that fits within GPU memory constraints.
Each angle block is marched sequentially, preventing the \gls{DGMOC} solver from exceeding memory limits.
During each angle block sweep, the following steps are performed: element-wise flat source moment calculation on the CPU, followed by transfer to the GPU for the entire domain, after which ray tracing, scalar flux restoration, and axial interface source calculation are performed on the GPU.
The ray tracing is similar to that of the planar \gls{MOC}, while several modifications are made to accommodate the sequential sweep in the other stages.
\begin{algorithm}[htbp]
  \caption{DGMOC sequential azimuthal angle sweep}
  \label{alg:DGMOC_sweep}
  \begin{algorithmic}[1]
    \State Compute element-wise flat source moments on the CPU and transfer them to the GPU for the entire domain
    \Statex \hspace{\algorithmicindent} \textit{(All subsequent operations are performed on the GPU)}    
    \For{each $angle\_block$ in azimuthal angle blocks}
    \For{$ward$ in \{upward, downward\}}
    \For{$iz$ in planes}
    \State Driven flux calculation for given $angle\_block$
    \State Ray tracing for given $angle\_block$
    \State Scalar flux restoration for $angle\_block$
    \State Axial interface source calculation for the next plane
    \EndFor
    \EndFor
    \EndFor
  \end{algorithmic}
\end{algorithm}

\FloatBarrier

%% file: sections/implementation/DFEM-SN.tex
\subsection{Memory Compression and Fine-grained GPU Parallelism in \gls{DFEMSN}}
\label{sec:DFEM-SN_implementation}

The \gls{DFEMSN} implementation in \gls{NuDEAL} focuses on optimizing angular flux storage and maximizing GPU parallelism.
Conventionally, an angle-by-angle sweep can reduce peak memory usage by allocating only the memory corresponding to the scalar flux.
However, this approach underutilizes GPU resources due to limited parallelism.
In contrast, solving all discrete angles simultaneously can significantly increase parallelism, but at the cost of requiring substantial memory to store angular flux over the entire domain.
To address this, \gls{NuDEAL} employs a compressed storage scheme for the angular fluxes, reducing memory usage without sacrificing accuracy.
This compression strategy leverages the fact that, during the sweep process, each angular element requires only angular flux information from its upstream neighbors.
That is, only the downstream neighbor elements need the angular flux of the already swept element.
By tracking the sweep order and element dependencies, \gls{NuDEAL} establishes a minimal set of angular flux storage locations, as enumerated in Algorithm~\ref{alg:DFEM_SN_sweep}.
Each angular element, denoted by $e$, is assigned a temporary storage slot at a certain sweep stage.
The slot is released only if all downstream elements $ne$ that depend on $e$ have been swept.
This slot-indexed angular flux buffer stores only the data required for the current sweep, thereby significantly reducing overall memory consumption.
A crucial aspect is that the slot index may be reused by angular elements that are spatially topologically distant, ensuring no conflicts during the sweep.
\begin{algorithm}[htbp]
  \caption{Establishing compressed angular flux storage scheme for DFEM-SN}
  \label{alg:DFEM_SN_sweep}
  \begin{algorithmic}[1]
    \State Given: boundary element set for a discrete angle
    \State Insert boundary element set into \texttt{Search} queue
    \While{\texttt{Search} queue is not empty}
    \For{each element \texttt{e} in \texttt{Search} queue}
    \For{each downstream element \texttt{ne} of element \texttt{e}}
    \If{all upstream elements of \texttt{ne} are visited}
    \State Push \texttt{ne} into \texttt{NextSearch} queue
    \Else
    \State Push \texttt{e} into \texttt{NextSearch} queue
    \EndIf
    \EndFor
    \EndFor
    \State Swap \texttt{Search} and \texttt{NextSearch} queues
    \EndWhile
  \end{algorithmic}
\end{algorithm}
\FloatBarrier

\begin{figure}[htbp]
  \centering
  \includegraphics[width=0.8\textwidth]{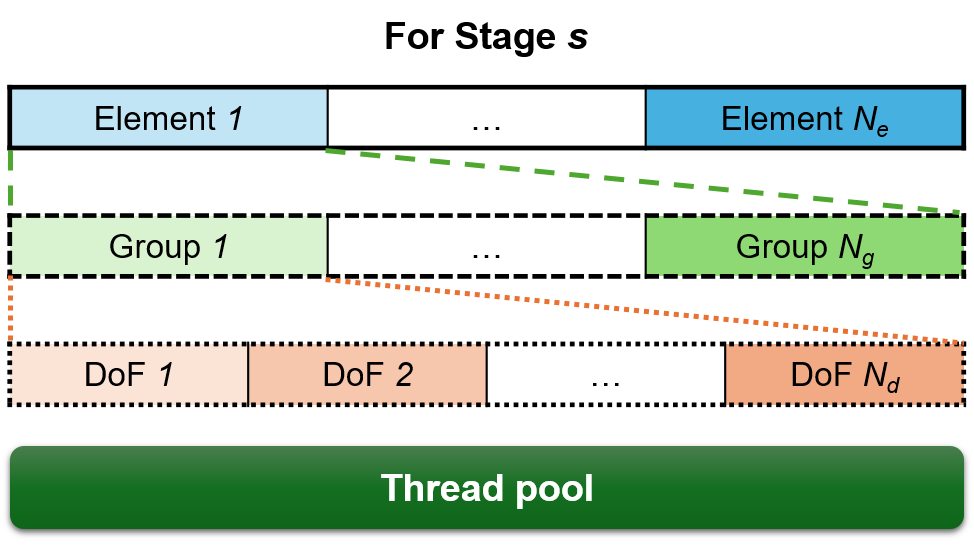}
  \caption{DoF-wide GPU parallelism for DFEM-SN sweep.}
  \label{fig:DFEMSNGPUSwepper}
\end{figure}

Besides, \gls{NuDEAL} employs fine-grained GPU parallelism by decomposing the sweep into smaller tasks that can be executed concurrently.
The overall solve procedure is composed of three GPU kernels:
\begin{enumerate}
  \item \texttt{InvestigateKernel}, which investigates the basic information, such as surface wind and neighboring elements, based on the current direction;
  \item \texttt{PrepareKernel}, which prepares the direction-dependent local matrix for each angular element; and
  \item \texttt{SweepKernel}, which performs the actual sweep by solving the local system for each element, energy group, and angle using the prepared matrices and the compressed angular flux storage.
\end{enumerate}
During the \texttt{SweepKernel}, \gls{NuDEAL} breaks down the sweep into \gls{DoF}-level tasks, instead of assigning one thread per angular element, as illustrated in Fig.~\ref{fig:DFEMSNGPUSwepper}.
Here, a unit \gls{DoF} is interpreted as a single angular flux moment and corresponds to a source moment and a rank in a local matrix.
Each thread is responsible for computing one \gls{DoF} of the angular flux for a given angular element, enabling a high degree of parallelism.
\FloatBarrier

%% file: sections/result.tex
\section{Numerical Results}
\label{sec:result}

\begin{figure}[htbp]
  \centering
  \begin{subfigure}[t]{0.4\linewidth}
    \centering
    \includegraphics[width=\linewidth]{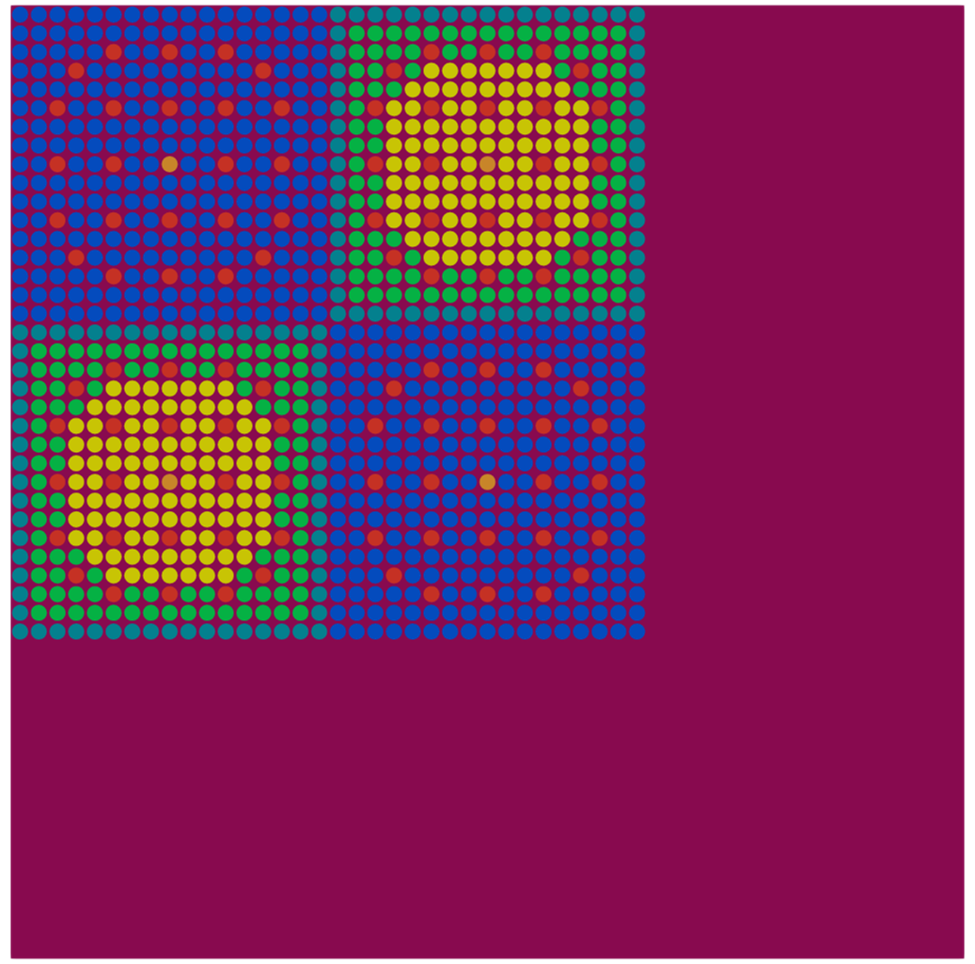}
    \caption{C5G7.}
    \label{fig:C5G72DConfigSub}
  \end{subfigure}
  \begin{subfigure}[t]{0.45\linewidth}
    \centering
    \includegraphics[width=\linewidth]{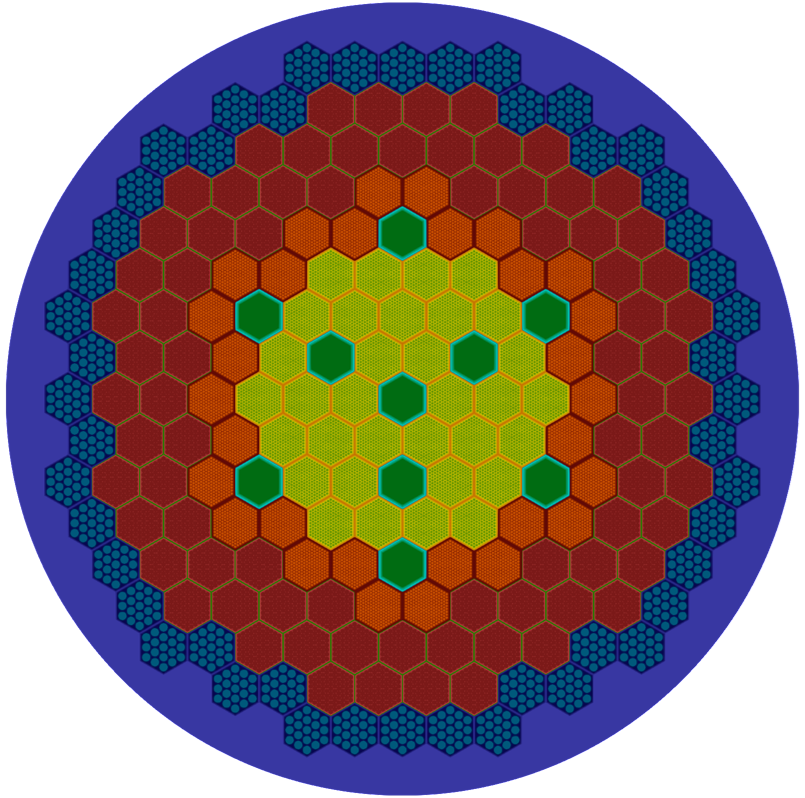}
    \caption{ABTR.}
    \label{fig:ABTR2DConfigSub}
  \end{subfigure}
  \vspace{1em}
  \begin{subfigure}[t]{0.45\linewidth}
    \centering
    \includegraphics[width=\linewidth]{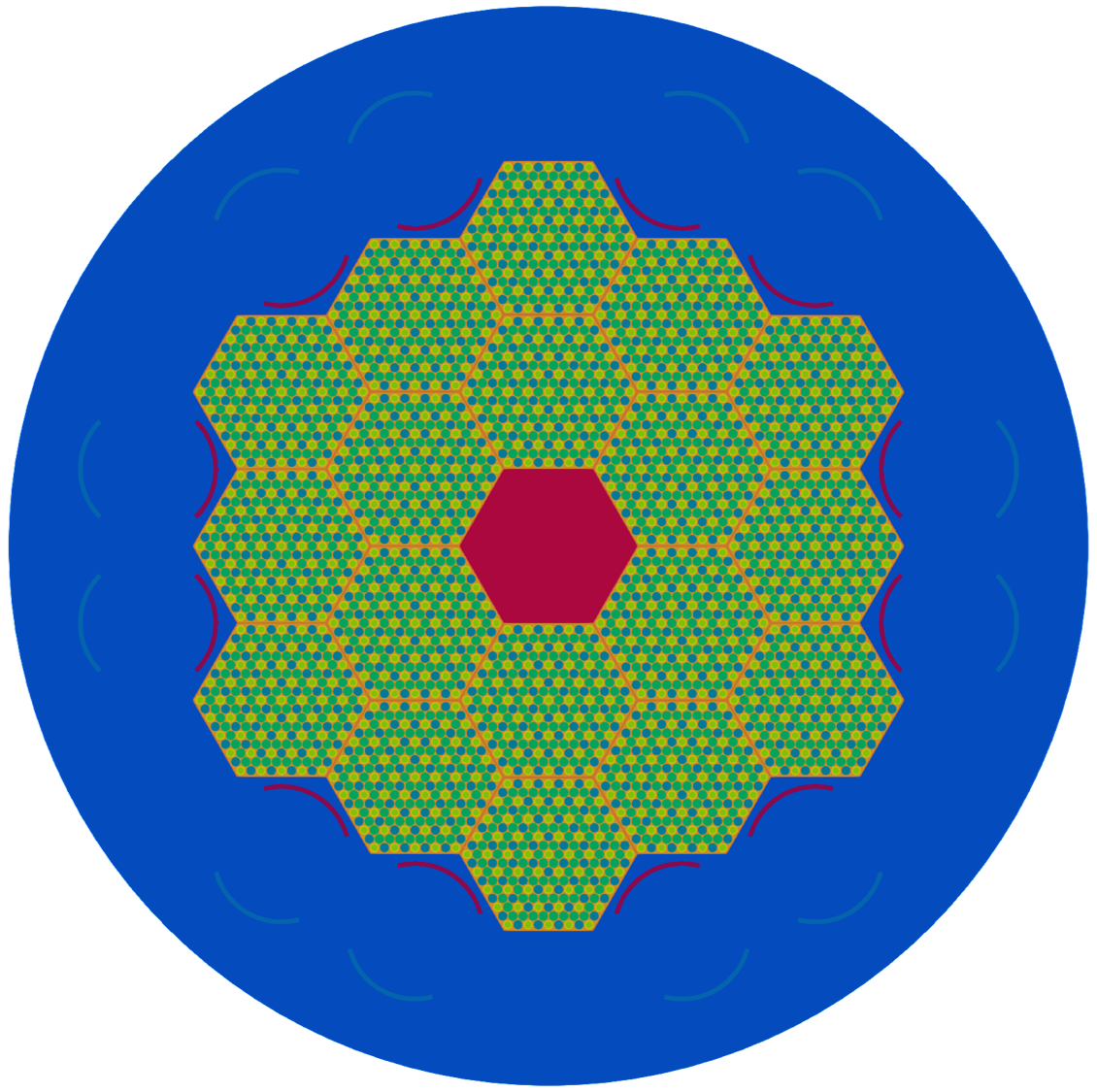}
    \caption{Empire.}
    \label{fig:Empire2DConfigSub}
  \end{subfigure}
  \begin{subfigure}[t]{0.45\linewidth}
    \centering
    \includegraphics[width=\linewidth]{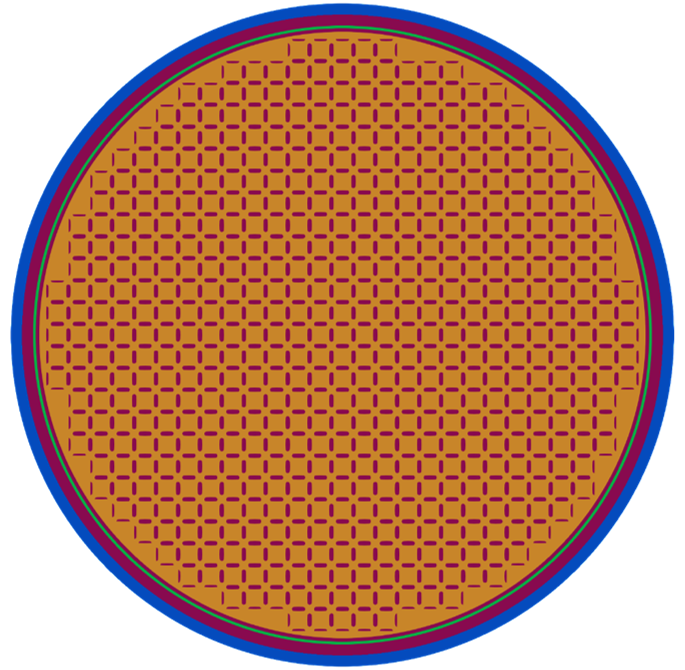}
    \caption{MSRE.}
    \label{fig:MSRE2DConfigSub}
  \end{subfigure}
  \caption{Benchmark problem 2D configurations.}
  \label{fig:benchmarkConfigs}
\end{figure}

This section presents the numerical results demonstrating the accuracy and performance of the implemented \gls{DGMOC}, \gls{DFEMSN}, and planar \gls{MOC}/\gls{HFEM} methods within the \gls{NuDEAL} framework.
Four benchmark problems were analyzed to verify and assess the developed solvers on unstructured meshes: C5G7~\cite{smith2005benchmark}, the \gls{ABTR}~\cite{kimBenchmarkSpecificationAdvanced2020}, Empire~\cite{dehartNEAMSReactorPhysics2020}, and the \gls{MSRE}~\cite{shenMOLTENSALTREACTOREXPERIMENTMSRE2006}.
All 2D configurations of the benchmark problems are illustrated in Fig.~\ref{fig:benchmarkConfigs}.
C5G7 serves as a structured reference benchmark representing a \gls{PWR}, whereas the other three correspond to advanced reactor concepts with unstructured geometries and heterogeneous spectra.
Except for C5G7, multigroup \glspl{XS} for these problems were generated using the GPU-based \gls{MC} code \gls{PRAGMA}~\cite{choiOptimizationNeutronTracking2021,imMultiphysicsAnalysisSystem2023a}, whose \gls{XS} generation workflow was previously verified for \gls{PWR} analyses~\cite{limPerformanceAnalysisTwostep2025}.
Then, the reference solution for those advanced reactors was obtained by performing multigroup \gls{MC} simulations using \gls{PRAGMA} unstructured mesh tracking capability~\cite{imMultiphysicsAnalysisSystem2023a} with sufficient particle histories to achieve low statistical uncertainties.

The C5G7 benchmark represents a quarter-core \gls{PWR} configuration composed of MOX and UO$_2$ assemblies surrounded by a radial reflector, as illustrated in Fig.~\ref{fig:C5G72DConfigSub}.
Three configurations, depending on the positions of the control rods, were analyzed to assess the effects of axial heterogeneity. \gls{ABTR} is a sodium-cooled fast reactor designed at \gls{ANL}, featuring a heterogeneous core layout of driver, control, and reflector assemblies arranged in a cylindrical vessel. Fig.~\ref{fig:ABTR2DConfigSub} illustrates the 2D core configuration used in this study. The original test and material assemblies are replaced with symmetric equivalents for modeling convenience. The Empire reactor is a conceptual heat pipe-cooled microreactor developed at \gls{LANL} that employs a monolithic core block in which fuel rods and heat pipes are arranged in a hexagonal lattice, as shown in Fig.~\ref{fig:Empire2DConfigSub}. The passive heat removal through embedded heat pipes introduces strong thermal coupling, while the compact heterogeneous geometry provides a challenging unstructured-mesh transport problem. \gls{MSRE} is a liquid-fueled thermal reactor operated at \gls{ORNL}, in which uranium tetrafluoride (UF$_4$) is dissolved in a LiF-BeF$_2$ salt mixture.
The core configuration, depicted in Fig.~\ref{fig:MSRE2DConfigSub}, consists of 100 fuel channels embedded in a graphite moderator matrix surrounded by a reflector.
\FloatBarrier

\input{sections/result/C5G7}
\input{sections/result/FastReactor}
\input{sections/result/ThermalReactor}
\input{sections/result/Discussion}

%% file: sections/result/C5G7.tex
\subsection{Verification on a Structured Reference Benchamrk: C5G7}
\label{sec:C5G7}

The C5G7 2D core is discretized into 189,584 quadrilateral elements.
The radial grid is refined to 283,584 elements to ensure a fair comparison with the legacy code when constructing 3D core problems.
Then, the core is constructed from 30 axial planes, yielding a total of 8.5 million hexahedral elements.
Table~\ref{tab:C5G7_2D_results} summarizes the eigenvalue and pin power errors of the \gls{MOC} and \gls{DFEMSN} methods for the 2D core problem.
Both methods demonstrate good agreement with the reference MCNP solutions~\cite{smith2005benchmark}, with eigenvalue errors within a few pcm and pin power errors below 1\%.
The \gls{DFEMSN} method shows slightly better accuracy in pin power distributions than the \gls{MOC} method.
This improved accuracy can be attributed to the linear source representation inherent to \gls{DFEMSN}, which provides a higher-order description of spatial source variation compared to the flat source approximation used in \gls{MOC}.
This advantage is expected to be more pronounced in the reflector region, where the mesh is relatively coarse, and the flat source approximation may be insufficient to accurately capture spatial variations.

\input{tables/01_C5G72DResult}
\FloatBarrier

For this problem, the \gls{DFEMSN} solver's performance is compared with the Griffin code, which employs the same methodology and is parallelized on CPUs.
The advantages of GPU acceleration and compressed angular flux storage are highlighted in Table~\ref{tab:vsGriffin}, which compares the problem complexity and runtime of the \gls{NuDEAL} simulation against Griffin's result~\cite{princePerformanceImprovementsGriffin2022}.
Although the \gls{NuDEAL} model has more elements and angles than the Griffin model, the \gls{NuDEAL} \gls{DFEMSN} implementation achieves comparable total computing time using a single GPU compared to Griffin running on 96 CPU cores.
Furthermore, the compressed angular flux storage scheme in \gls{NuDEAL}, denoted as \textit{Concurrent}, significantly reduces GPU memory requirements and reduces total computing time compared to the traditional angle-by-angle sweep, denoted as \textit{Sequential}.
\input{tables/02_vsGriffin}
\FloatBarrier

To evaluate the performance of the developed \gls{MOC} solver on unstructured meshes, the GPU ray tracing module in nTRACER is used as a reference structured-grid \gls{MOC} solver~\cite{choiPracticalAccelerationDirect2021}.
Since the original 2D core problem is not big enough to effectively demonstrate the performance difference, a larger problem is created by duplicating the active core 4 times in both $x$ and $y$ directions, resulting in a $8 \times 8$ fuel assembly array.
Table~\ref{tab:vsnTRACER} compares the problem complexity and runtime between \gls{NuDEAL} and nTRACER for the enlarged problem.
\texttt{Ratio} denotes the ratio of \gls{NuDEAL} to nTRACER values for each metric.
Although the \gls{NuDEAL} model has more elements and segments than the nTRACER model due to the unstructured mesh representation, the \gls{NuDEAL} \gls{MOC} implementation achieves a significantly faster single sweep time and total computing time compared to nTRACER.
Although it is hard to quantify, this performance gap is attributed to several implementation differences between the two codes:
\begin{enumerate}
  \item C++/CUDA for \gls{NuDEAL} vs.\ CUDA Fortran for nTRACER,
  \item 128-bit memory alignment in \gls{NuDEAL} vs.\ 32-bit alignment in nTRACER.
\end{enumerate}
Both are related because CUDA Fortran does not support 128-bit memory alignment.
These implementation choices lead to better memory access patterns and improved computational efficiency in \gls{NuDEAL}.
\input{tables/03_vsnTRACER}
\FloatBarrier

For the 3D core problem, planar \gls{MOC}/\gls{HFEM} and \gls{DGMOC} methods are employed, whereas \gls{DFEMSN} is excluded due to its excessive memory requirements, which are prohibitive for a single GPU.
Table~\ref{tab:C5G7_3D_results} summarizes the eigenvalue and pin power errors of the planar \gls{MOC}/\gls{HFEM} and \gls{DGMOC} methods.
Diffusion and SP3 approximations are considered for \gls{HFEM}.
\gls{HFEM} takes about 30 seconds and 4 minutes for diffusion and SP3 approximations, respectively, while 7 minutes is consumed for the \gls{DGMOC} calculation.
Accuracy improvements are observed when moving from the diffusion to the SP3 approximation in \gls{HFEM}.
The \gls{DGMOC} method achieves the best accuracy among the three methods, but at the cost of increased computational time.
Since \gls{HFEM} utilizes the subplane scheme~\cite{choAxialSPNRadial2007,zhangQuadraticAxialExpansion2020} to take advantage of a thick plane for \gls{MOC} calculations, the computing cost of \gls{HFEM} is much cheaper than that of \gls{DGMOC}.
\input{tables/04_C5G73DResult}
\FloatBarrier

The performance of the \gls{NuDEAL} \gls{DGMOC} solver is compared with the PROTEUS-MOC code, a legacy code employing the same methodology on CPUs.
Table~\ref{tab:vsPROTEUS} presents the problem complexity and runtime comparison between \gls{NuDEAL} and PROTEUS-MOC for the unrodded configuration~\cite{zhangQuadraticAxialExpansion2020}.
\gls{NuDEAL} employs a single GPU to complete the simulation faster than PROTEUS-MOC.
Especially, the azimuthal angle sequential sweeping in \gls{NuDEAL} does not degrade the performance of the \gls{DGMOC} solver, as illustrated in Table~\ref{tab:DGMOC_angle_seq}.
A parallel azimuthal-angle sweep reduces the single-transport-sweep time by only 3.0\% compared to sequential execution.
This is because the \gls{DGMOC} method involves a sufficiently large number of computational tasks per azimuthal angle, enabling effective utilization of GPU resources even in sequential angle sweeping.
Given that memory pressure increases significantly with larger-angle block sizes, the sequential-sweeping strategy is a favorable choice for the \gls{DGMOC} implementation on GPUs.
\input{tables/05_vsPROTEUS}
\input{tables/06_DGMOC_angle_seq}
\FloatBarrier

%% file: tables/01_C5G72DResult.tex
\begin{table}[htbp]
  \centering
  \caption{Calculation results of the C5G7 2D core.}
  \label{tab:C5G7_2D_results}
  \begin{tabular}{|c|c|cccc|}
    \hline
    \multirow{2}{*}{Method} & \multirow{2}{*}{\makecell[c]{Eigenvalue                                                           \\error (pcm)}} & \multicolumn{4}{c|}{Pin power error (\%)}                         \\ \cline{3-6}
                            &                                         & Max.$^{\,a}$ & Avg.$^{\,b}$ & RMS$^{\,c}$ & MRE$^{\,d}$ \\ \hline
    MOC                     & -3                                      & 0.829        & 0.148        & 0.205       & 0.120       \\
    DFEM-SN                 & 1                                       & 0.667        & 0.127        & 0.169       & 0.104       \\ \hline
  \end{tabular}

  \vspace{2mm}
  {\footnotesize
    \raggedright
    $^{a}$Maximum pin power error.\\
    $^{b}$Average pin power error.\\
    $^{c}$Root-mean-square pin power error.\\
    $^{d}$Mean relative pin power error.\\}
\end{table}

%% file: tables/02_vsGriffin.tex
\begin{table}[htbp]
  \centering
  \caption{C5G7 2D core DFEM-SN calculation breakdown against Griffin.}
  \label{tab:vsGriffin}
  \begin{tabular}{|c|c|>{\centering\arraybackslash}p{2.2cm}|>{\centering\arraybackslash}p{2.2cm}|}
    \hline
    Code                        & Griffin                                     & \multicolumn{2}{c|}{NuDEAL}                          \\
    \hline
    Angle treatment in sweeping & --                                          & \textit{Sequential}            & \textit{Concurrent} \\ \hline
    Processor                   & \makecell[c]{96 CPU cores                                                                          \\Intel Xeon Platinum}
                                & \multicolumn{2}{c|}{\makecell[c]{Single GPU                                                        \\NVIDIA RTX 4070 Ti}} \\ \hline
    \# of Elements              & 87{,}108                                    & \multicolumn{2}{c|}{189{,}584}                       \\ \hline
    \# of Angles                & 192                                         & \multicolumn{2}{c|}{256}                             \\ \hline
    Total Angular flux DoF (GB) & 2.6                                         & \multicolumn{2}{c|}{10}                              \\ \hline
    \# of Transport sweeps      & 17                                          & \multicolumn{2}{c|}{32}                              \\ \hline
    Total computing time (s)    & 53                                          & 148                            & 26                  \\ \hline
    \makecell[c]{GPU memory for                                                                                                      \\angular flux storage (MB)} & -- & 40 & 23 \\ \hline
  \end{tabular}
\end{table}

%% file: tables/03_vsnTRACER.tex
\begin{table}[htbp]
  \centering
  \caption{Problem complexity and runtime ratio between NuDEAL and nTRACER.}
  \label{tab:vsnTRACER}
  \begin{tabular}{|c|c|c|c|c|}
    \hline
    Code    & \# of Elements & \# of Segments & Single sweep time (s) & Computing time (s) \\ \hline
    NuDEAL  & 2.7M           & 307M           & 0.5                   & 32                 \\ \hline
    nTRACER & 1.5M           & 250M           & 0.9                   & 46                 \\ \hline
    (Ratio) & 1.8            & 1.2            & 0.56                  & 0.70               \\ \hline
  \end{tabular}
\end{table}

%% file: tables/04_C5G73DResult.tex
\begin{table}[htbp]
  \centering
  \caption{Calculation results of the C5G7 3D core.}
  \label{tab:C5G7_3D_results}
  \begin{tabular}{llrrrrr}
    \toprule
    Method & Config.  & $\Delta k_{\text{eff}}^{\,a}$ (pcm) & \multicolumn{4}{c}{Max. pin power error (\%)}                                  \\
    \cmidrule(lr){4-7}
           &          &                                     & Slice 1                                       & Slice 2 & Slice 3 & Int.$^{b}$ \\
    \midrule
    \multirow{3}{*}{DGMOC}
           & Unrodded & 11                                  & 0.953                                         & 1.129   & 1.566   & 0.980      \\
           & Rodded A & 13                                  & 1.098                                         & 1.167   & 1.319   & 0.900      \\
           & Rodded B & -7                                  & 1.047                                         & 1.158   & 1.774   & 1.131      \\
    \midrule
    \multirow{3}{*}{MOC/HFEM Diffusion}
           & Unrodded & -60                                 & 1.426                                         & 1.130   & 1.807   & 0.874      \\
           & Rodded A & -55                                 & 1.280                                         & 1.147   & 1.636   & 0.762      \\
           & Rodded B & -114                                & 1.404                                         & 1.295   & 2.519   & 1.206      \\
    \midrule
    \multirow{3}{*}{MOC/HFEM SP$_3$}
           & Unrodded & 30                                  & 1.150                                         & 1.079   & 1.506   & 0.778      \\
           & Rodded A & 45                                  & 1.188                                         & 1.061   & 1.575   & 0.669      \\
           & Rodded B & 39                                  & 1.191                                         & 0.916   & 1.657   & 0.848      \\
    \bottomrule
  \end{tabular}

  \vspace{2mm}
  {\footnotesize
    \raggedright
    $^{a}$Eigenvalue error.\\
    $^{b}$Error in axially integrated pin power.\\}
\end{table}

%% file: tables/05_vsPROTEUS.tex
\begin{table}[htbp]
  \centering
  \caption{Performance comparison between PROTEUS-MOC and NuDEAL.}
  \label{tab:vsPROTEUS}
  \begin{tabular}{lcc}
    \toprule
    Code                                                                  & PROTEUS-MOC & NuDEAL \\
    \midrule
    Processor                                                             &
    \begin{tabular}[c]{@{}c@{}}
      480 CPU cores \\ Blue Gene/P$^{*}$
    \end{tabular} &
    \begin{tabular}[c]{@{}c@{}}
      Single GPU \\ NVIDIA RTX 2080 Ti
    \end{tabular}                                                              \\
    \midrule
    \# of Elements                                                        & (Unknown)   &
    \begin{tabular}[c]{@{}c@{}}
      8{,}507{,}520 \\ (283{,}584 $\times$ 30)
    \end{tabular}                                                      \\
    \midrule
    \# of Transport sweeps                                                & 28          & 36     \\
    Single transport sweep time (s)                                       & 8.05        & 5.25   \\
    \midrule
    Total computing time (s)                                              & 566         & 412    \\
    \bottomrule
  \end{tabular}

  \vspace{2mm}
  {\footnotesize
    \raggedright
    $^{*}$Processor information for PROTEUS-MOC is inferred from the number of cores reported in~\cite{zhangQuadraticAxialExpansion2020}, together with the known hardware at \gls{ANL} at the time~\cite{marin-laflecheDevelopmentStatusPROTEUSMOC2012}.\\}
\end{table}

%% file: tables/06_DGMOC_angle_seq.tex
\begin{table}[htbp]
  \centering
  \caption{Efficiency of sequential azimuthal angle sweep in DGMOC.}
  \label{tab:DGMOC_angle_seq}
  \begin{tabular}{|c|c|c|c|}
    \hline
    Size of angle block &
    \makecell[c]{Single transport               \\ sweep time (s)} &
    Single MOC time (s) &
    \makecell[c]{GPU                            \\ memory pressure (MB)}                    \\
    \hline
    $2^{a}$             & 5.25 & 10.5 & 2{,}589 \\
    4                   & 5.19 & 9.96 & 2{,}893 \\
    8                   & 5.10 & 9.50 & 3{,}499 \\
    16                  & 5.10 & 9.30 & 4{,}709 \\
    32                  & 5.09 & 9.21 & 7{,}133 \\
    $64^{b}$            & N.A. & N.A. & N.A.    \\
    \hline
  \end{tabular}

  \vspace{2mm}
  {\footnotesize
    \raggedright
    $^{a}$Fully sequential execution.\\
    $^{b}$Fully concurrent execution but out of memory.\\}
\end{table}

%% file: sections/result/FastReactor.tex
\subsection{Application to Fast Reactors: \gls{ABTR} and Empire}
\label{sec:FastReactor}

\begin{figure}[htbp]
  \centering
  \includegraphics[width=0.9\textwidth]{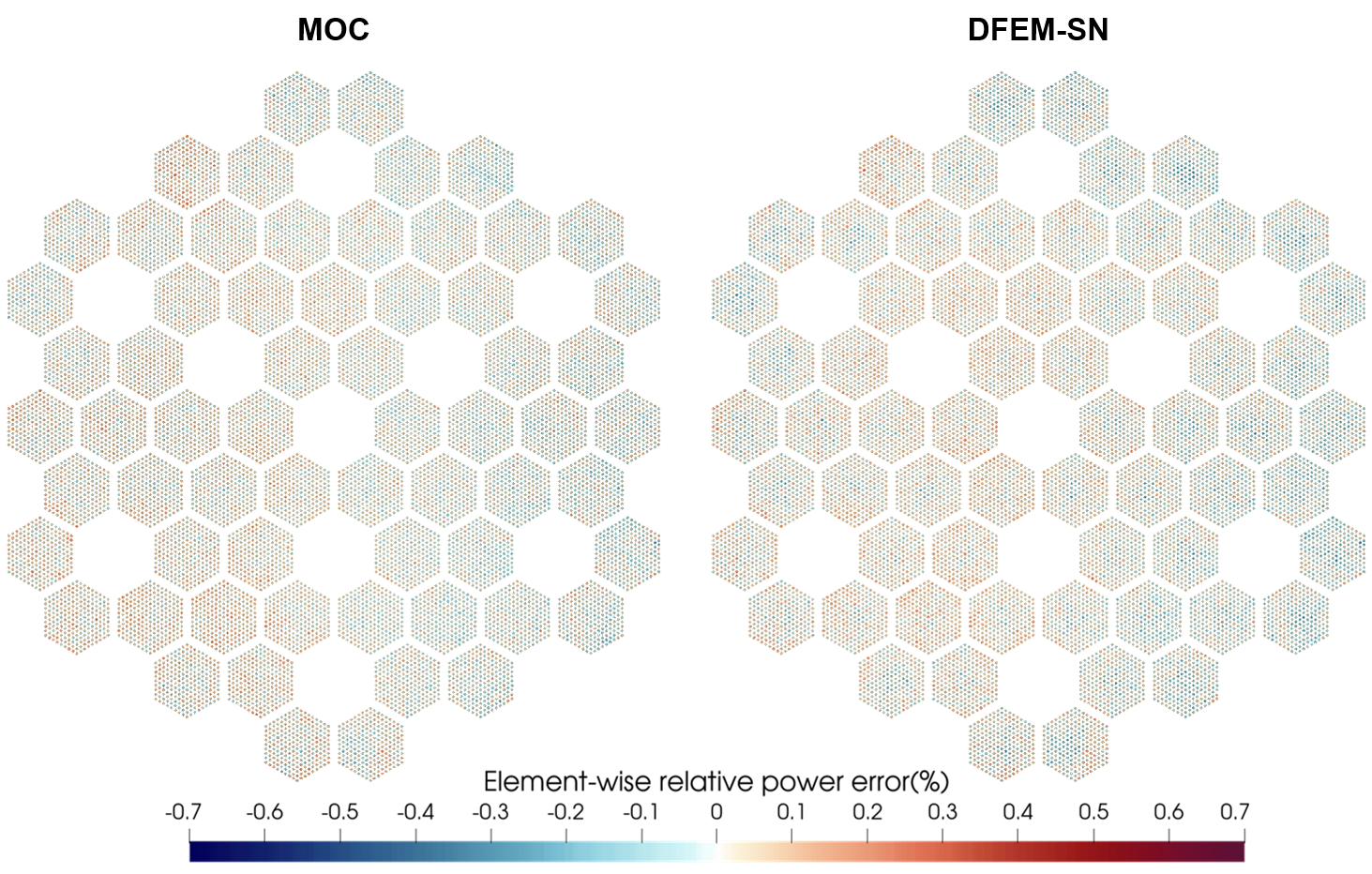}
  \caption{ABTR 2D core element-wise relative power error distribution.}
  \label{fig:ABTR2DElemPowerDiff}
\end{figure}

The 2D core mesh for the \gls{ABTR} benchmark consists of approximately 1.6 million elements.
Table~\ref{tab:ABTR_2D_results} summarizes the eigenvalue and pin power errors of the \gls{MOC} and \gls{DFEMSN} methods for this 2D core problem.
Both methods demonstrate good agreement with the reference \gls{MC} solution, with eigenvalue errors around 30 pcm and pin power errors below 0.3\%.
\gls{MOC} completes the calculation in less than a minute, whereas \gls{DFEMSN} requires 4 minutes.
Since Fig.~\ref{fig:ABTR2DElemPowerDiff} indicates that no distinguishable difference is found in the element-wise power error distribution of \gls{MOC} and \gls{DFEMSN}, \gls{MOC} is considered a more efficient choice for this problem.
\input{tables/07_ABTR2DCore}
\FloatBarrier

\begin{figure}[htbp]
  \centering
  \includegraphics[width=0.8\textwidth]{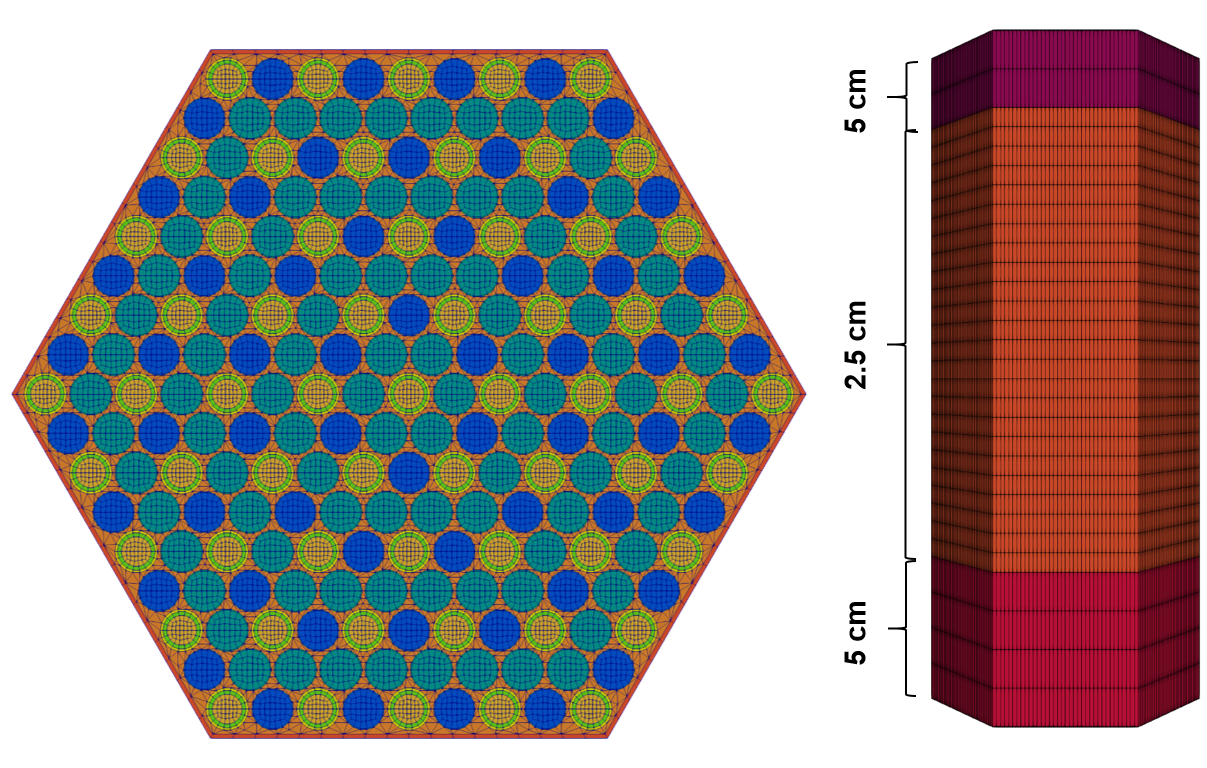}
  \caption{Empire fuel assembly mesh in the radial (left) and axial (right) views.}
  \label{fig:EmpireAssemblyMesh}
\end{figure}

For the Empire benchmark, a fuel assembly and a core configuration are solved.
The assembly is meshed with approximately 42,000 elements in 2D, and the mesh is extruded to create a 3D problem with 30 slabs, as shown in Fig.~\ref{fig:EmpireAssemblyMesh}.
Table~\ref{tab:EmpireFuelAssembly2D} and Table~\ref{tab:EmpireFuelAssembly3D} present the eigenvalue results for the 2D and 3D Empire fuel assembly problems, respectively.
In the 2D case, \gls{DFEMSN} shows better accuracy than \gls{MOC} in terms of eigenvalue.
The \gls{MOC}/\gls{HFEM}--$SP_{3}$ method is evaluated for coarse-mesh generation in 3D using various stencil grid sizes.
The stencil grid size does not impact the eigenvalue results, with errors remaining around 344 pcm.
\gls{HFEM} shows a slight improvement in accuracy over \gls{DGMOC}, while \gls{DFEMSN} achieves the best accuracy.
\input{tables/08_EmpireFuelAssembly2D}
\input{tables/09_EmpireFuelAssembly3D}
\FloatBarrier

\begin{figure}[htbp]
  \centering
  \includegraphics[width=0.8\textwidth]{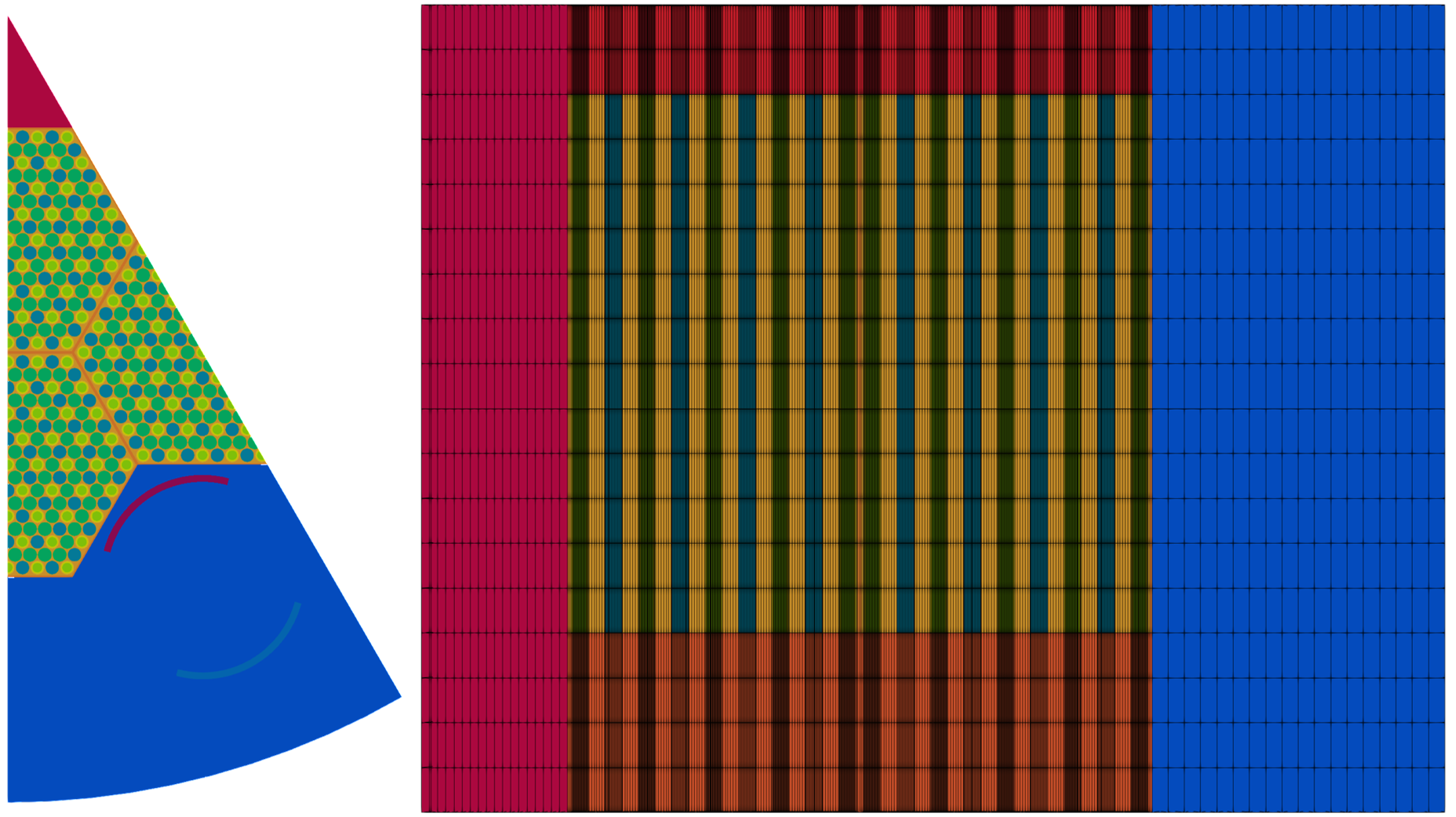}
  \caption{Empire 1/12 core configuration in the radial (left) and axial (right) views.}
  \label{fig:EmpireCoreMesh}
\end{figure}

\begin{figure}[htbp]
  \centering
  \includegraphics[width=0.8\textwidth]{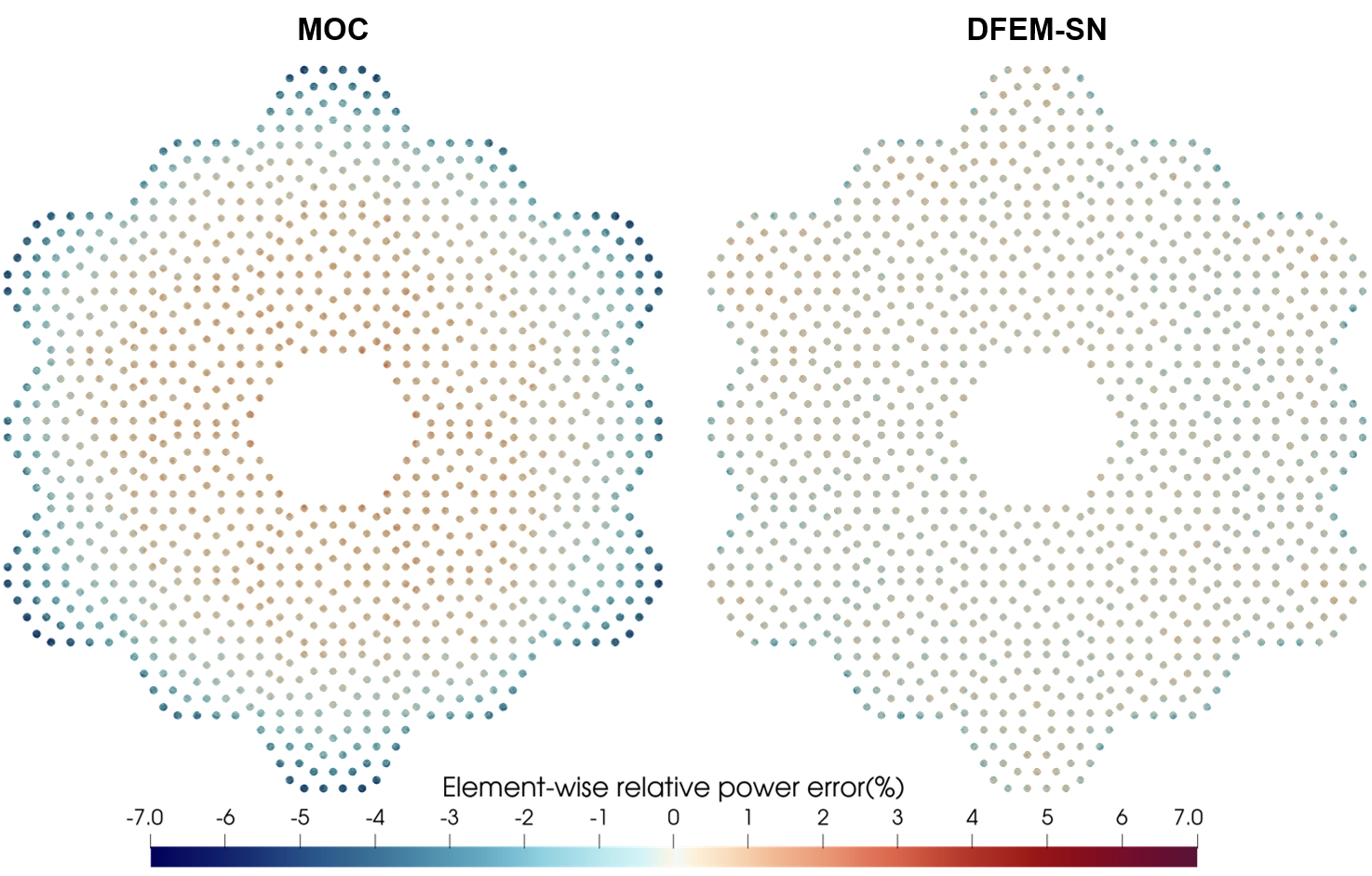}
  \caption{Empire 2D core element-wise relative power error distribution.}
  \label{fig:Empire2DCoreElemPowerDiff}
\end{figure}

\begin{figure}[htbp]
  \centering
  \includegraphics[width=0.8\textwidth]{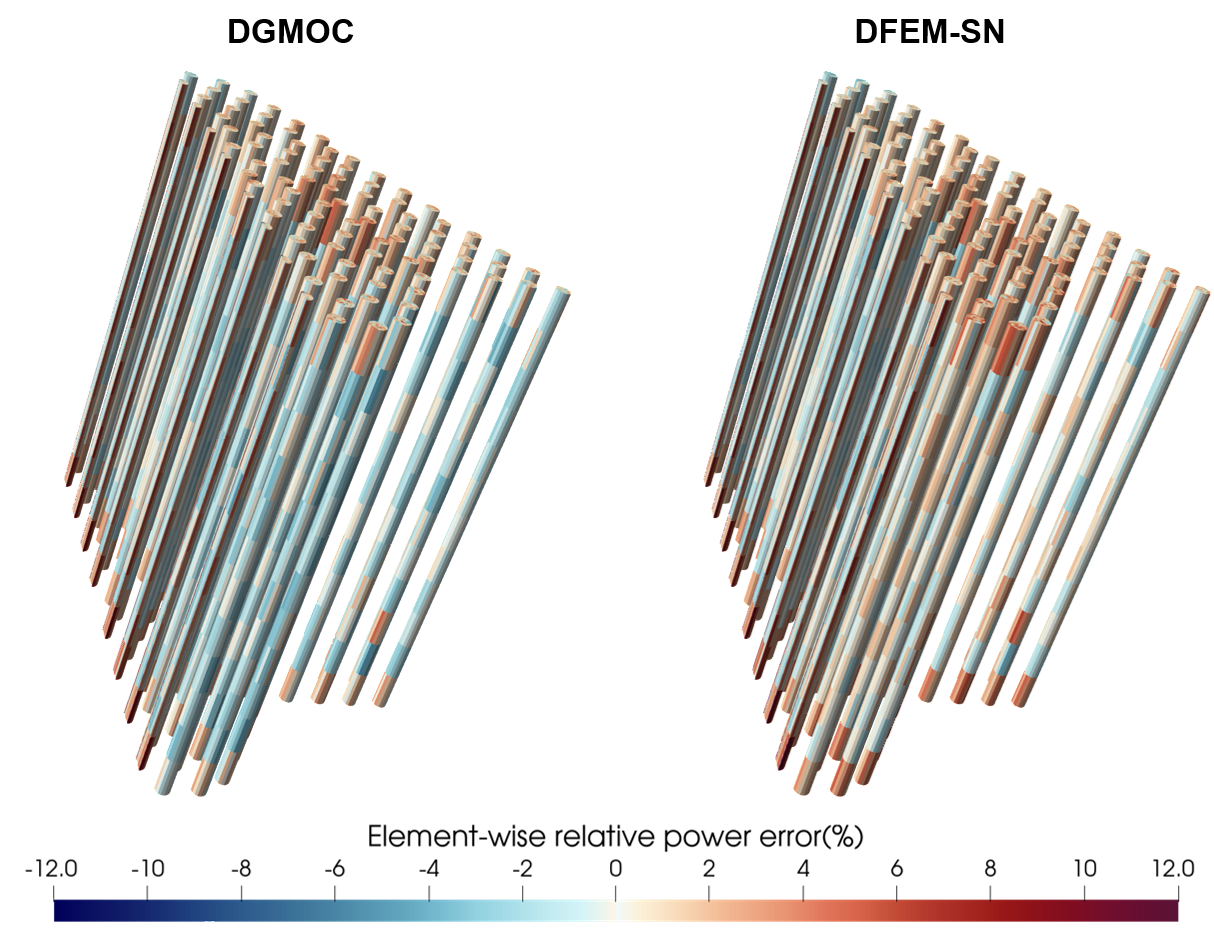}
  \caption{Empire 3D 1/12 core element-wise relative power error distribution.}
  \label{fig:Empire3DCoreElemPowerDiff}
\end{figure}

The Empire 2D core is meshed with approximately 360,000 elements.
For the 3D problem, 1/12 symmetry is utilized due to the \gls{DFEMSN}'s heavy memory requirement, resulting in 590 thousand elements, which is illustrated in Fig.~\ref{fig:EmpireCoreMesh}.
Table~\ref{tab:Empire_2D_core} summarizes the 2D core calculation results.
\gls{DFEMSN} demonstrates superior accuracy in element and pin power distributions at the cost of larger computing time than \gls{MOC}.
Fig.~\ref{fig:Empire2DCoreElemPowerDiff} indicates that the \gls{DFEMSN} power distribution error is more uniform and agrees better with the reference solution than \gls{MOC}.
Similarly, \gls{DFEMSN} outperforms \gls{DGMOC} in the 3D core calculation in terms of the pin power distribution, as shown in Table~\ref{tab:Empire_3D_core}.
However, the computational cost is more pronounced, with \gls{DFEMSN} requiring 4,882 seconds compared to \gls{DGMOC}'s 290 seconds.
The maximum power errors appear around the top and bottom reflector regions, as shown in Fig.~\ref{fig:Empire3DCoreElemPowerDiff}.
Notably, \gls{HFEM} failed to converge for the 3D core problem.
It is observed that the air hole located in the center of the core, reserved for control rods, causes convergence difficulties for \gls{HFEM}, which is sensitive to near-void regions.
\input{tables/10_EmpireCore2D}
\input{tables/11_EmpireCore3D}
\FloatBarrier

%% file: tables/07_ABTR2DCore.tex
\begin{table}[htbp]
  \centering
  \caption{Calculation results of the ABTR 2D core.}
  \label{tab:ABTR_2D_results}
  \begin{tabular}{lccccccc}
    \toprule
    \multirow{2}{*}{Solver}                      & \multirow{2}{*}{$\Delta k$ (pcm)} &
    \multicolumn{2}{c}{Element power error (\%)} &
    \multicolumn{2}{c}{Pin power error (\%)}     &
    \multirow{2}{*}{\makecell[c]{Computing                                                                                                                                                          \\time (minute)}} \\
    \cmidrule(lr){3-4} \cmidrule(lr){5-6}
                                                 &                                   & $\Delta P_{\text{MAX}}$ & $\Delta P_{\text{RMS}}$ & $\Delta P_{\text{MAX}}$ & $\Delta P_{\text{RMS}}$ &      \\
    \midrule
    DFEM-SN                                      & $-34$                             & 0.606                   & 0.121                   & 0.263                   & 0.062                   & 4    \\
    MOC                                          & $-32$                             & 0.596                   & 0.116                   & 0.233                   & 0.052                   & $<1$ \\
    \bottomrule
  \end{tabular}

  \vspace{2mm}
  \raggedright
  \footnotesize
  $\Delta k$: Eigenvalue difference against multigroup MC calculation, 1.27539 ($\pm$1).\\
  $\Delta P_{\text{MAX}}$: Maximum relative power error.\\
  $\Delta P_{\text{RMS}}$: RMS relative power error.\\
  \# of elements: 1,608,458.
\end{table}

%% file: tables/08_EmpireFuelAssembly2D.tex
\begin{table}[htbp]
  \centering
  \caption{Calculation results of the Empire 2D fuel assembly.}
  \label{tab:EmpireFuelAssembly2D}
  \begin{tabular}{lcc}
    \toprule
    \multicolumn{2}{c}{Multigroup MC reference (std, pcm)} & 1.28232 ($\pm 1$)         \\
    \midrule
    \multirow{2}{*}{$\Delta k$ (pcm)}                      & DFEM-SN           & $27$  \\
                                                           & MOC               & $150$ \\
    \bottomrule
  \end{tabular}
\end{table}

%% file: tables/09_EmpireFuelAssembly3D.tex
\begin{table}[htbp]
  \centering
  \caption{Calculation results of the Empire 3D fuel assembly.}
  \label{tab:EmpireFuelAssembly3D}
  \begin{tabular}{lccc}
    \toprule
    \multicolumn{3}{c}{Multigroup MC reference (std, pcm)} & 1.20582 ($\pm 1$)                                    \\
    \midrule
    \multirow{6}{*}{$\Delta k$ (pcm)}                      & \multicolumn{2}{c}{DFEM-SN} & 294                    \\
                                                           & \multicolumn{2}{c}{DGMOC}   & 395                    \\
                                                           & \multirow{4}{*}{MOC/HFEM}   & $1 \times 1^{*}$ & 345 \\
                                                           &                             & $2 \times 2$     & 345 \\
                                                           &                             & $4 \times 4$     & 343 \\
                                                           &                             & $8 \times 8$     & 344 \\
    \bottomrule
  \end{tabular}

  \vspace{2mm}
  \raggedright
  \footnotesize
  $^{*}$Stencil grid used for the coarse mesh employed in the HFEM calculations.\\
\end{table}

%% file: tables/10_EmpireCore2D.tex
\begin{table}[htbp]
  \centering
  \caption{Empire 2D core calculation summary.}
  \label{tab:Empire_2D_core}
  \begin{tabular}{lccccccc}
    \toprule
    \multirow{2}{*}{Solver}                      & \multirow{2}{*}{$\Delta k$ (pcm)} &
    \multicolumn{2}{c}{Element power error (\%)} &
    \multicolumn{2}{c}{Pin power error (\%)}     &
    \multirow{2}{*}{\makecell[c]{Computing                                                                                                                                                         \\time (s)}} \\
    \cmidrule(lr){3-4} \cmidrule(lr){5-6}
                                                 &                                   & $\Delta P_{\text{MAX}}$ & $\Delta P_{\text{RMS}}$ & $\Delta P_{\text{MAX}}$ & $\Delta P_{\text{RMS}}$ &     \\
    \midrule
    DFEM-SN                                      & 34                                & 3.319                   & 0.515                   & 1.483                   & 0.289                   & 100 \\
    MOC                                          & 45                                & 6.390                   & 1.325                   & 4.816                   & 1.258                   & 10  \\
    \bottomrule
  \end{tabular}

  \vspace{2mm}
  \raggedright
  \footnotesize
  $\Delta k$: Eigenvalue difference against multigroup MC calculation, 1.12012 ($\pm$2).\\
  \# of elements: 359,454.\\
\end{table}

%% file: tables/11_EmpireCore3D.tex
\begin{table}[htbp]
  \centering
  \caption{Empire 3D 1/12 core calculation summary.}
  \label{tab:Empire_3D_core}
  \begin{tabular}{lccccccc}
    \toprule
    \multirow{2}{*}{Solver}                      & \multirow{2}{*}{$\Delta k$ (pcm)} &
    \multicolumn{2}{c}{Element power error (\%)} &
    \multicolumn{2}{c}{Pin power error (\%)}     &
    \multirow{2}{*}{\makecell[c]{Computing                                                                                                                                                           \\time (s)}} \\
    \cmidrule(lr){3-4} \cmidrule(lr){5-6}
                                                 &                                   & $\Delta P_{\text{MAX}}$ & $\Delta P_{\text{RMS}}$ & $\Delta P_{\text{MAX}}$ & $\Delta P_{\text{RMS}}$ &       \\
    \midrule
    DFEM-SN                                      & 153                               & 11.399                  & 1.596                   & 0.859                   & 0.249                   & 4,882 \\
    DGMOC                                        & 216                               & 9.501                   & 1.572                   & 1.121                   & 0.375                   & 290   \\
    \bottomrule
  \end{tabular}

  \vspace{2mm}
  \raggedright
  \footnotesize
  $\Delta k$: Eigenvalue difference against multigroup MC calculation, 1.12043 ($\pm$2).\\
  \# of elements: 587,628.\\
\end{table}

%% file: sections/result/ThermalReactor.tex
\subsection{Application to Advanced Thermal Reactor: \gls{MSRE}}
\label{sec:ThermalReactor}

\begin{figure}[htbp]
  \centering
  \begin{subfigure}[t]{0.45\linewidth}
    \centering
    \includegraphics[width=0.5\linewidth]{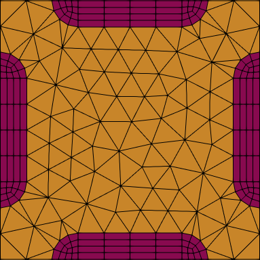}
    \caption{Base.}
    \label{fig:MSREChannelModelBase}
  \end{subfigure}
  \begin{subfigure}[t]{0.45\linewidth}
    \centering
    \includegraphics[width=0.5\linewidth]{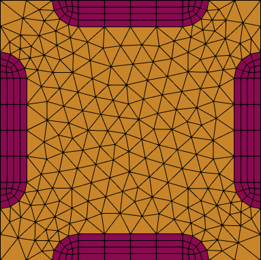}
    \caption{Refined.}
    \label{fig:MSREChannelModelRefined}
  \end{subfigure}
  \caption{MSRE 2D mesh for fuel channel lattice.}
  \label{fig:MSREChannelModels}
\end{figure}

\begin{figure}[htbp]
  \centering
  \begin{subfigure}[t]{0.45\linewidth}
    \centering
    \includegraphics[width=\linewidth]{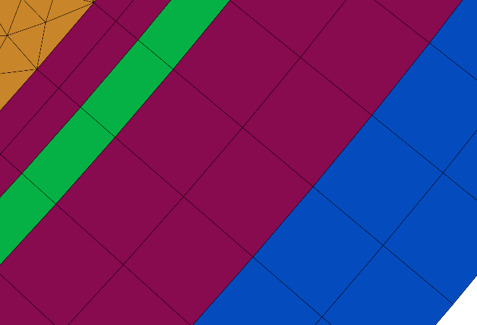}
    \caption{Base.}
    \label{fig:MSREStructureModelBase}
  \end{subfigure}
  \begin{subfigure}[t]{0.45\linewidth}
    \centering
    \includegraphics[width=\linewidth]{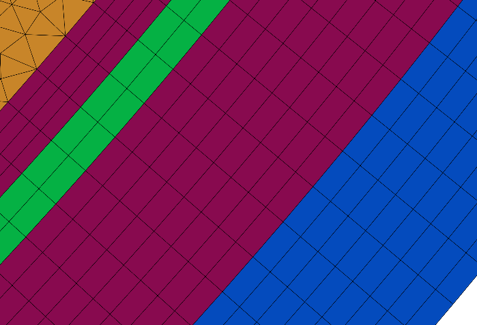}
    \caption{Refined 1.}
    \label{fig:MSREStructureModel1}
  \end{subfigure}
  \vspace{1em}
  \begin{subfigure}[t]{0.45\linewidth}
    \centering
    \includegraphics[width=\linewidth]{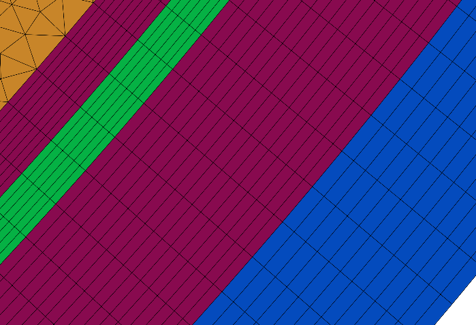}
    \caption{Refined 2.}
    \label{fig:MSREStructureModel2}
  \end{subfigure}
  \begin{subfigure}[t]{0.45\linewidth}
    \centering
    \includegraphics[width=\linewidth]{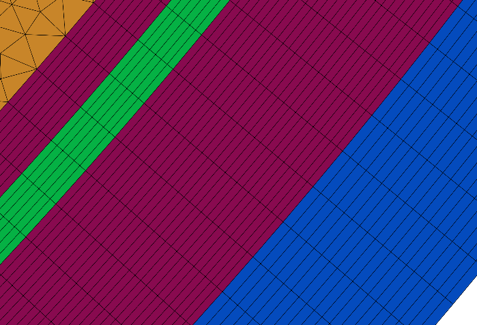}
    \caption{Refined 3.}
    \label{fig:MSREStructureModel3}
  \end{subfigure}
  \caption{MSRE 2D mesh for outer structure.}
  \label{fig:MSREStructureModels}
\end{figure}

Fig.~\ref{fig:MSREChannelModels} and Fig.~\ref{fig:MSREStructureModels} illustrate the unstructured meshes for the \gls{MSRE} 2D core problem, having various refinement levels.
Among these meshes, the base mesh shown in Fig.~\ref{fig:MSREChannelModelBase} and Fig.~\ref{fig:MSREStructureModelBase}, the least dense one, is selected for the baseline calculation.
Table~\ref{tab:MSRE2DCoreBaseLine} summarizes the eigenvalue results for the \gls{MSRE} 2D core problem.
\gls{DFEMSN} achieves an eigenvalue error of 22 pcm, while \gls{MOC} shows a larger error of -918 pcm.
These results indicate that \gls{DFEMSN} achieves excellent accuracy for the \gls{MSRE} core problem, even with the baseline mesh, which can be attributed to its linear representation of the element-wise source term, allowing improved accuracy on relatively coarse meshes.
\input{tables/12_MSRE2DCoreBaseLine}
\FloatBarrier

\begin{figure}
  \centering
  \includegraphics[width=\textwidth]{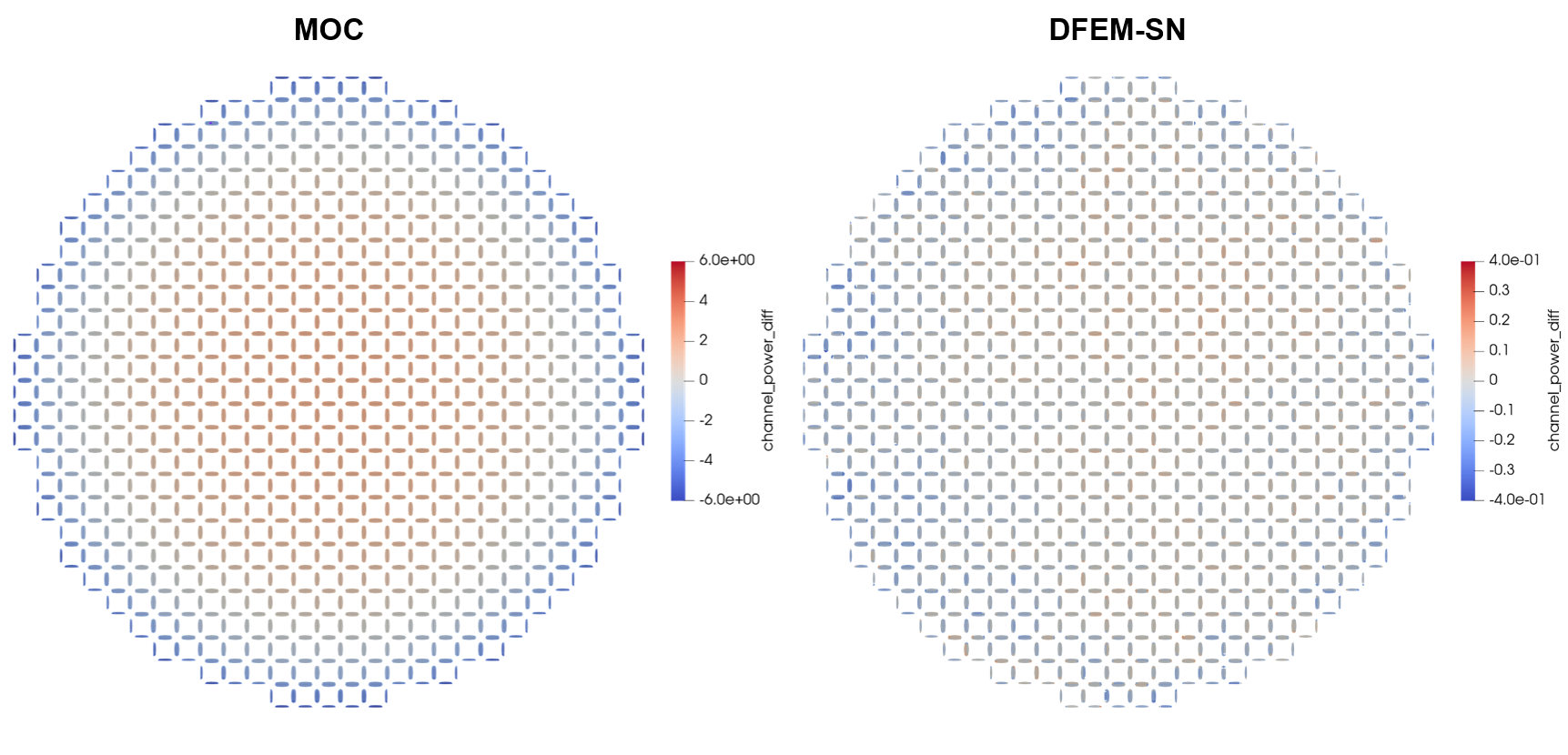}
  \caption{MSRE 2D core channel-wise relative power error distribution for the baseline mesh.}
  \label{fig:MSRE2DCoreChannelPowerDiffBaseline}
\end{figure}

\begin{figure}
  \centering
  \includegraphics[width=\textwidth]{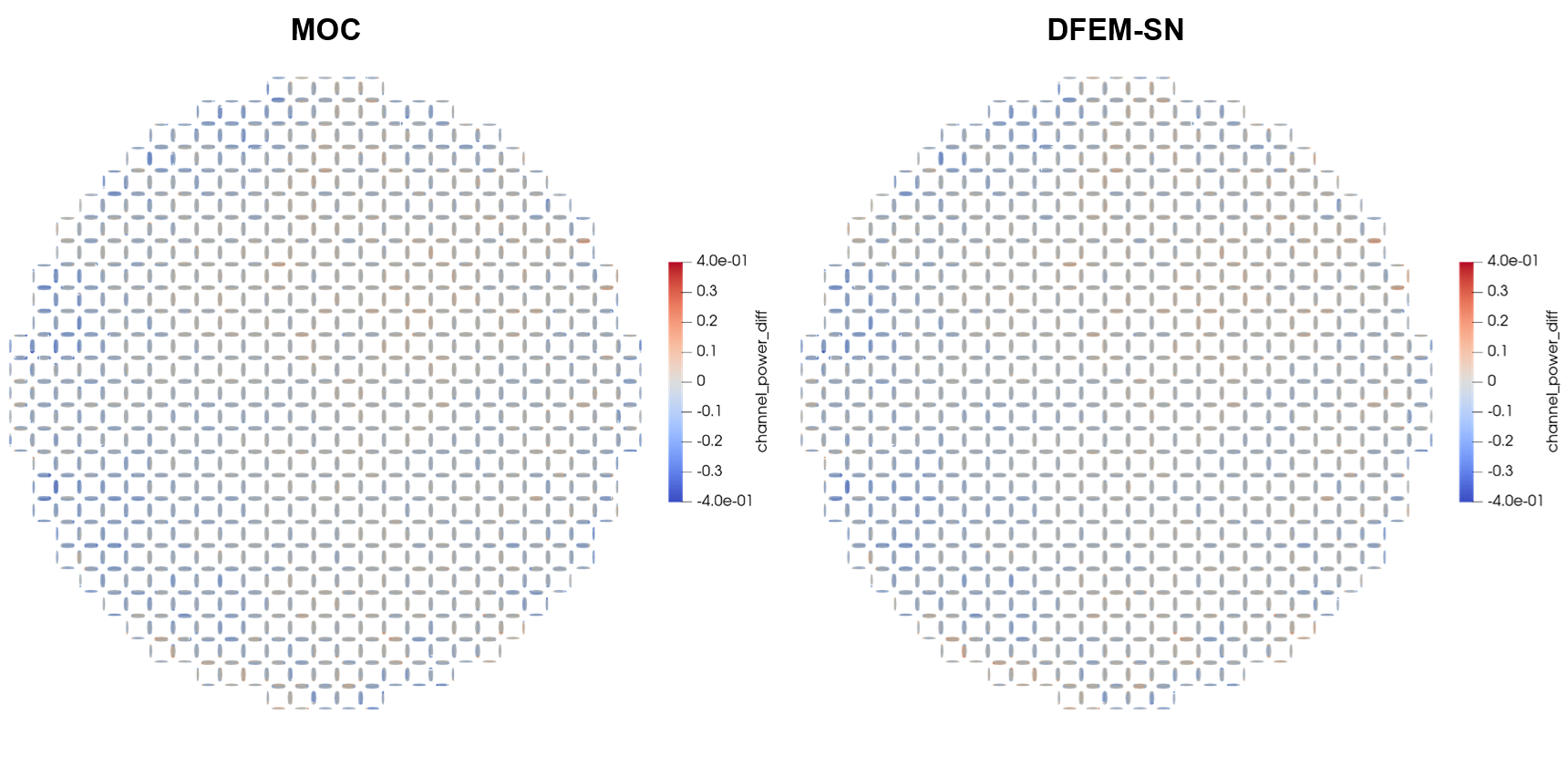}
  \caption{MSRE 2D core channel-wise relative power error distribution for the most refined mesh.}
  \label{fig:MSRE2DCoreChannelPowerDiffRefined}
\end{figure}

Further investigations were conducted to assess the impact of mesh refinement on the \gls{MOC} solver's eigenvalue results.
Table~\ref{tab:MSRE2DCoreMeshRefinement} presents the eigenvalue results from \gls{MOC} calculations on meshes with varying resolutions.
As the mesh is refined, the eigenvalue error decreases significantly from -918 pcm in the base mesh to -62 pcm in the most refined mesh (Case 4).
In particular, refining the outer-structure mesh notably improves accuracy, as observed in Case 1.
This error behavior is also reflected in the power distribution results.
Fig.~\ref{fig:MSRE2DCoreChannelPowerDiffBaseline} and Fig.~\ref{fig:MSRE2DCoreChannelPowerDiffRefined} illustrate the channel-wise relative power error distributions for the baseline and most refined meshes, respectively.
In the baseline results, a larger power error of 6\% is observed in the \gls{MOC} calculation, whereas \gls{DFEMSN} maintains errors below 0.4\%.
\gls{MOC} can achieve accuracy comparable to \gls{DFEMSN} only by using the most refined mesh, thereby increasing the computational cost.
This sensitivity stems from the linear source approximation inherent in \gls{DFEMSN}, which effectively captures flux variations within each element, even on coarser meshes.
\input{tables/13_MSRE2DCoreMeshRefinement}
\FloatBarrier

\begin{figure}[htbp]
  \centering
  \includegraphics[width=0.95\textwidth]{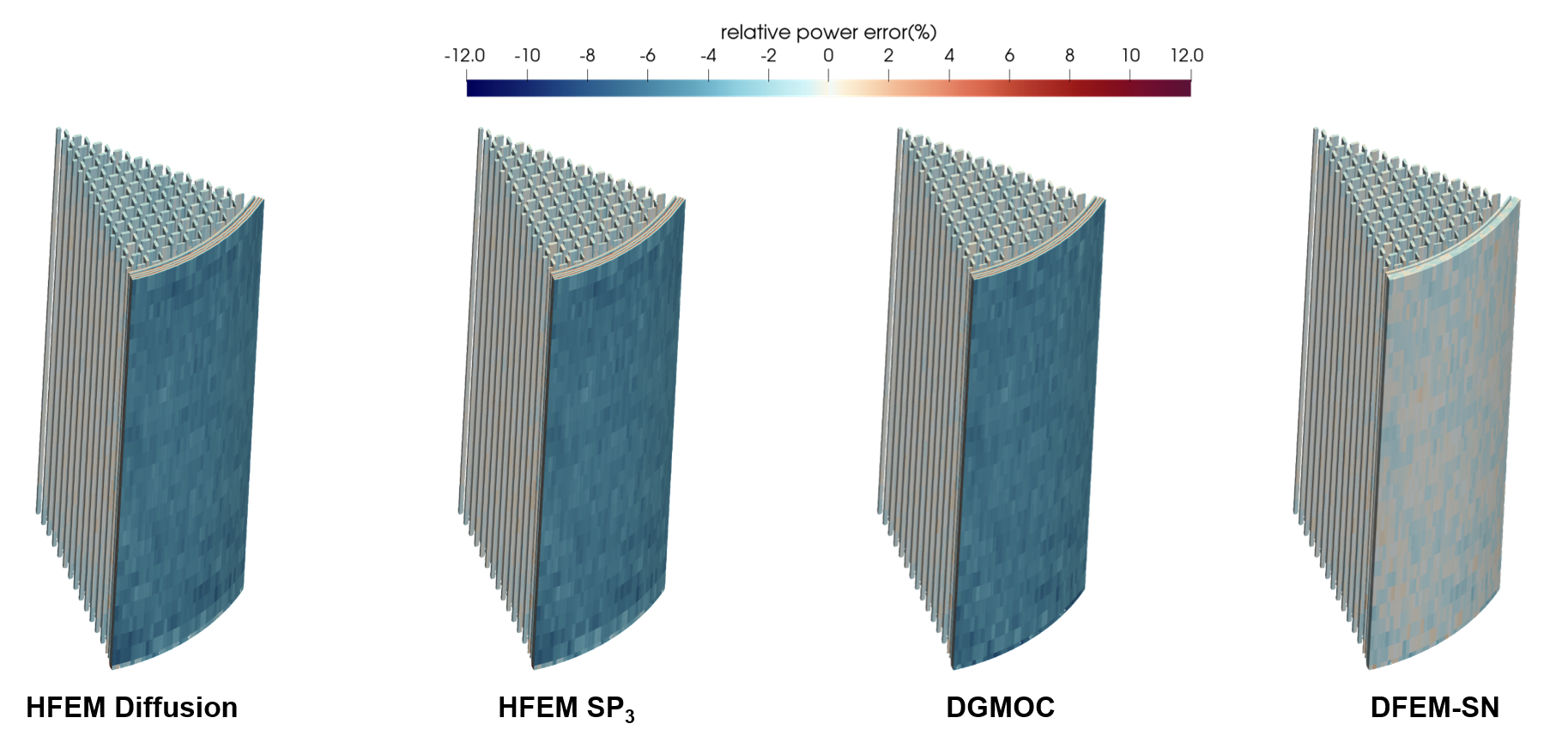}
  \caption{MSRE 3D octant core element-wise power error distribution.}
  \label{fig:MSRE3DCoreElemPowerDiff}
\end{figure}

The 3D configuration of the \gls{MSRE} core is modeled using the octant symmetry due to the computational resource limitations.
Table~\ref{tab:MSRE3D_octant_core} summarizes the eigenvalue and power distribution results for the \gls{MSRE} 3D octant core problem.
\gls{DFEMSN} achieves the best accuracy with an eigenvalue error of 6 pcm and a maximum power error of 4.7\%, at the cost of the longest computing time over an hour.
\gls{MOC} with both diffusion and SP3 \gls{HFEM} also provide reasonable accuracy with eigenvalue errors of -46 pcm and -23 pcm, respectively, and both take less than a minute of computing time.
\gls{DGMOC} is positioned between \gls{MOC}/\gls{HFEM} and \gls{DFEMSN} in terms of accuracy and computing time, showing a trade-off between the two.
Fig.~\ref{fig:MSRE3DCoreElemPowerDiff} illustrates the element-wise power error distribution for the \gls{MSRE} 3D octant core problem.
It is observed that \gls{MOC}-based solvers tend to underestimate the power distribution at the peripheral regions of the core.
Since the peripheral power is approximately 1\% of the average core power, this results in a larger relative error in those regions.
Nevertheless, \gls{DFEMSN} maintains a more accurate power distribution even at the core's periphery.
\input{tables/14_MSRE3DCore}
\FloatBarrier

%% file: tables/12_MSRE2DCoreBaseLine.tex
\begin{table}[htbp]
  \centering
  \caption{MSRE 2D core baseline calculation results.}
  \label{tab:MSRE2DCoreBaseLine}
  \begin{tabular}{lcc}
    \toprule
    \multicolumn{2}{c}{Multigroup MC reference (std, pcm)} & 1.15900 ($\pm 1$)        \\
    \midrule
    \multirow{2}{*}{$\Delta k$ (pcm)}                      & DFEM-SN           & 22   \\
                                                           & MOC               & -918 \\
    \bottomrule
  \end{tabular}
\end{table}

%% file: tables/13_MSRE2DCoreMeshRefinement.tex
\begin{table}[htbp]
  \centering
  \caption{MSRE 2D core MOC calculation results on refined meshes with varying resolution.}
  \label{tab:MSRE2DCoreMeshRefinement}
  \begin{tabularx}{\textwidth}{|c|c|*{5}{>{\centering\arraybackslash}X|}}
    \hline
    \multicolumn{2}{|c|}{Case}                 & Base            & 1                         & 2         & 3         & 4                              \\
    \hline
    \multirow{2}{*}{Model}                     & Fuel channel    & \multicolumn{4}{c|}{Base} & Refined                                                \\
    \cline{2-7}
                                               & Outer structure & Base                      & Refined 1 & Refined 2 & \multicolumn{2}{c|}{Refined 3} \\
    \hline
    \multicolumn{2}{|c|}{\# of elements}       & 190k            & 210k                      & 230k      & 240k      & 360k                           \\
    \hline
    \multicolumn{2}{|c|}{MOC $\Delta k$ (pcm)} & -918            & -215                      & -140      & -124      & -62                            \\
    \hline
  \end{tabularx}
\end{table}

%% file: tables/14_MSRE3DCore.tex
\begin{table}[htbp]
  \centering
  \caption{MSRE 3D octant core calculation results.}
  \label{tab:MSRE3D_octant_core}
  \begin{tabularx}{\textwidth}{|c|*{4}{>{\centering\arraybackslash}X|}}
    \hline
    \multirow{2}{*}{Solver}             & \multicolumn{2}{c|}{MOC/HFEM} & \multirow{2}{*}{DGMOC} & \multirow{2}{*}{DFEM-SN}         \\
    \cline{2-3}
                                        & Diffusion                     & SP$_3$                 &                          &       \\
    \hline
    \# of Elements                      & \multicolumn{3}{c|}{1.7M}     & 0.9M                                                      \\
    \hline
    $\Delta k$ (pcm)                    & -46                           & -23                    & -90                      & 6     \\
    \hline
    $\Delta P_{\text{MAX}}$ (\%) $^{a}$ & 11.0                          & 10.0                   & 8.4                      & 4.7   \\
    $(P>P_{avg})$ $^{b}$                & 3.3                           & 2.8                    & 2.5                      & 2.7   \\
    \hline
    $\Delta P_{\text{RMS}}$ (\%)        & 1.7                           & 1.6                    & 1.6                      & 0.9   \\
    $(P>P_{avg})$                       & 0.42                          & 0.38                   & 0.37                     & 0.37  \\
    \hline
    Computing time (s)                  & 23                            & 34                     & 436                      & 4,685 \\
    \hline
  \end{tabularx}

  \vspace{2mm}
  \raggedright
  \footnotesize
  $\Delta P$: Relative power error against the multigroup MC calculation.\\
  $a$: Error for the entire domain.\\
  $b$: Error for elements having above-average power (1.0).\\
\end{table}

%% file: sections/result/Discussion.tex
\subsection{Discussion}
\label{sec:discussion}

The verification and application results presented in the previous subsections highlight the complementary roles of the three deterministic transport solvers developed in \gls{NuDEAL}.
Each method exhibits distinct strengths in accuracy, efficiency, and applicability, depending on the problem's spectral characteristics and geometric complexity.
Across all benchmarks, the \gls{DFEMSN} method consistently achieved the highest accuracy.
Its finite element nature effectively captured intra-element flux gradients even on relatively coarse unstructured meshes, as clearly observed in the \gls{MSRE} case.
However, this accuracy came at the cost of increased memory demand and longer runtime, particularly for 3D problems.
\gls{DGMOC} provided a balanced compromise between accuracy and efficiency.
Its axial \gls{DG} expansion enabled improved axial flux representation while maintaining the computational efficiency of ray tracing.
The sequential azimuthal angle sweep achieved high GPU utilization while keeping memory usage manageable.
The planar \gls{MOC}/\gls{HFEM} approach demonstrated excellent computational efficiency by coupling fine-mesh radial transport with a coarse-mesh 3D nodal solver.
On the other hand, its accuracy is limited because no rigorous closure or equivalence principle is directly applied between \gls{MOC} and \gls{HFEM}.
As seen in the Empire 3D core, near-void regions can challenge \gls{HFEM} because the total \gls{XS} appears in the denominator, highlighting the nature of the second-order transport equation.

Mesh quality affects solver accuracy, depending on the spectral regime.
In fast-spectrum systems such as \gls{ABTR} and Empire, further mesh refinement does not improve the solution accuracy of any method, as angular resolution and treatment of streaming effects dominate solution accuracy.
On the contrary, as observed in the \gls{MSRE} case, spatial discretization and scattering representation become more influential in the thermal spectrum system, where \gls{DFEMSN}’s linear source property provides a distinct advantage.
This result also indicates that mesh refinement in this system can substantially reduce \gls{MOC} errors, reducing eigenvalue discrepancies by nearly 900 pcm.
This observation confirms that spatial resolution rather than algorithmic formulation often limits accuracy in diffusion-dominated configurations.

The GPU implementations in \gls{NuDEAL} demonstrated that fine-grained parallelism and memory-aware design can make deterministic whole-core calculations on unstructured meshes practically feasible.
The 128-bit aligned memory structure, sequential azimuthal angle sweep, and compressed angular-flux storage substantially reduced latency and GPU memory pressure, enabling single- and gaming-GPU simulations comparable to legacy CPU clusters, such as PROTEUS-MOC on 480 cores and Griffin on 96 cores.
These results suggest that carefully tailored data-parallel algorithms can close the gap between deterministic and \gls{MC} methods in both fidelity and scalability, even on the complex geometries represented as unstructured meshes.

%% file: sections/conclusion.tex
\section{Conclusion and Future Work}
\label{sec:conclusion}

This study developed and verified three complementary deterministic neutron transport solvers, \gls{DGMOC}, \gls{DFEMSN}, and a planar \gls{MOC}/\gls{HFEM}, within a unified GPU-accelerated framework, \gls{NuDEAL}, for high-fidelity analysis of advanced reactor systems on unstructured meshes.
All the methods were formulated consistently from the multigroup transport equation and implemented with memory-aware GPU parallelization strategies to achieve both numerical rigor and computational practicality.

Comprehensive benchmark evaluations demonstrated the accuracy and performance of the developed solvers across structured and unstructured geometries.
In the C5G7 benchmark, both \gls{MOC} and \gls{DFEMSN} reproduced the reference results within a few pcm in eigenvalues and below 1\% in pin-power error, thereby verifying the correctness of the formulations.
The \gls{DGMOC} method reproduced the PROTEUS-MOC results on a single GPU and demonstrated that the sequential azimuthal-angle sweep can maintain high throughput with minimal overhead.
The GPU-based \gls{DFEMSN} solver achieved performance comparable to that of Griffin on 96 CPU cores, while the \gls{MOC} implementation surpassed nTRACER GPU-accelerated ray tracing in speed due to its optimized 128-bit memory alignment.

Applications to advanced reactor problems confirmed the robustness of the solvers under diverse spectral and geometric conditions.
For fast-spectrum systems, all methods predicted eigenvalues within 400 pcm of the multigroup \gls{MC} results, with \gls{DFEMSN} achieving the best accuracy and \gls{MOC} the shortest runtime.
For the thermal-spectrum \gls{MSRE}, \gls{DFEMSN} again provided the highest fidelity, while \gls{MOC} showed strong convergence toward the same accuracy as the mesh was refined.
These results indicate that deterministic GPU solvers can now deliver both precision and performance sufficient for whole-core unstructured-mesh simulations that were previously accessible only to large CPU clusters.

Further developments and research directions are envisioned to enhance the capabilities and applicability of the \gls{NuDEAL} framework.
Regarding the \gls{MOC}/\gls{HFEM} method, devising a loose consistency scheme to couple \gls{MOC} and \gls{HFEM} is a priority to improve accuracy by applying a closure or equivalence principle between them.
Additionally, implementing a near-void treatment for \gls{HFEM} using a mixed-hybrid formulation~\cite{lewisMuchAdoNothing2004} may alleviate numerical instability in low-density regions.
For \gls{DGMOC} and \gls{DFEMSN}, elaborating parallelization strategies and optimizing data structures can further improve GPU utilization and reduce memory footprint.
On the physics side, integrating the \gls{XS} library generation and transplanting the resonance treatment modules will enhance the solvers' overall capabilities.
Finally, implementing domain decomposition and load-balancing schemes will enable efficient handling of large-scale problems on multi-GPU clusters.
In particular, the \gls{DGMOC} and \gls{DFEMSN} methods are expected to benefit significantly from domain decomposition, making them increasingly feasible and scalable in environments equipped with multiple GPUs.